\documentclass[11pt,a4paper]{article}
\pdfoutput=1 % if your are submitting a pdflatex (i.e. if you have
\usepackage{jheppub}
\usepackage[svgnames,x11names]{xcolor} % Required to specify font color

\usepackage[utf8]{inputenc}
\usepackage{dsfont}
\usepackage{amsfonts}
\usepackage{amsmath}
\usepackage{amsthm}
\usepackage{graphicx}
\usepackage{wrapfig}
\usepackage{calligra}
\newcommand{\ii}{\textrm{i}}
\newcommand{\dd}{\textrm{d}}
\newcommand{\e}{\textrm{e}}

\newcommand{\del}{\partial}

\newcommand{\bb}{> \hspace{-5pt}>}

\newcommand{\ga}{\alpha}
\newcommand{\gb}{\beta}
\renewcommand{\gg}{\gamma}
\newcommand{\gd}{\delta}
\newcommand{\gep}{\epsilon}
\newcommand{\gz}{\zeta}

\newcommand{\gth}{\theta}

\newcommand{\gl}{\lambda}

\newcommand{\gs}{\sigma}

\newcommand{\gw}{\omega}

\newcommand{\Gg}{\Gamma}

\newcommand{\Wbar}{\overline{W}}

\newcommand{\cC}{\mathcal C}

\newcommand{\cN}{\mathcal N}
\newcommand{\cO}{\mathcal O}
\newcommand{\cP}{\mathcal P}

\newcommand{\cR}{\mathcal R}
\newcommand{\cS}{\mathcal S}

\newcommand{\cW}{\mathcal W}

\newcommand{\Tr}{\mbox{Tr}}
\newcommand{\tr}{\mbox{Tr}}
\newcommand{\be}{\begin{equation}}
\newcommand{\bea}{\begin{eqnarray}}
\newcommand{\ee}{\end{equation}}
\newcommand{\eea}{\end{eqnarray}}
\renewcommand{\d}{\partial}

\newcommand{\half}{\frac{1}{2}}

\newcommand{\ov}[1]{\frac{1}{#1}}

\title{Notes on BPS Wilson Loops and the Cusp Anomalous Dimension in ABJM theory}

\author{Daniele Marmiroli}
\affiliation{Nordita, KTH Royal Institute of Technology and Stockholm University, Roslagstullsbacken 23,\\ SE-10691 Stockholm, Sweden} 
\emailAdd{daniele.marmiroli@nordita.org}
\vspace{60mm}

\abstract{We introduce a new purely bosonic, $\frac{1}{6}$ BPS Wilson loop for ABJM theory on $S^3$ that couples scalar fields to a latitude at an angle $\theta$ on $S^2\in\mathbb{C}P^3$. Through localization of this operator, we  relate the expansion of the cusp anomalous dimension at small cusp angles to the logarithmic derivative of the ABJM Wilson loop. This defines, non-perturbatively in the 't~Hooft coupling, the bremsstrahlung function $B(\lambda)$ describing in three dimensions the soft radiation of a $W$-boson undergoing a sudden change in trajectory. We compare our results for $B(\lambda)$ to the known weak/strong  coupling expansions of the function $h(\lambda)$ that enters integrability. At weak coupling we precisely match the previously known two-loop result. At strong coupling we find agreement at leading order in $\sqrt{\lambda}$, but a  mismatch of the constant coefficient. We comment on the striking similarity that we observe between these two, in principle, unrelated functions. }


\begin{document}

\maketitle

\newpage

\section{Overview}

In supersymmetric gauge theories the logarithmic divergences due to the emission of soft fields by a $W$-boson  undergoing a sudden change in its trajectory are encoded in the cusp anomalous dimension $\Gamma(\gl,\gamma,\gth)$ defined through \cite{Korchemsky:1987wg}

$$
\log\left<\cW(\cC)\right> \sim - \Gamma(\gl,\gamma,\gth) \log\frac{L}{\gep}
$$

$\cW(\cC)$ is the Wilson loop operator coupled to a contour $\cC$ developing a cusp of angle $\gamma$, $\gth$ is some internal angle, $L$ and $\gep$ are infrared and ultraviolet cutoffs and $\gl$ is the 't~Hooft coupling constant. $\Gamma(\gl,\gamma,\gth)$ is also related to the anomalous dimension of twist-two operators with large spin \cite{Craigie:1980qs,Balitsky:1987bk,Korchemsky:1992xv}, therefore its name. It follows that such function constitutes a fundamental element for the understanding of the gauge theory both perturbatively, in the weakly and strongly coupled regimes, and non-perturbatively, by means of localization and integrability. Analytically continuing the internal angle $\gth\to\ii\vartheta$ it is nontheless possible to perform Bethe-Salpeter resummation in the $\vartheta\to\infty$ limit \cite{Correa:2012nk,Bykov:2012sc,Henn:2012qz}, which provides a further genuine field theoretic result in the strongly coupled region.
Moreover, a non-perturbative expression for the small $\gamma$, small $\gth$ behaviour  $\Gamma(\gl,\gamma,\gth)$ in $\cN=4$ SYM theory was derived in \cite{Correa:2012at} by relating the latter to the expectation value, exactly computed by means of localization, of the circular Wilson loop \cite{Pestun:2007rz}. Such expression was also checked against a TBA computation \cite{Correa:2012hh,Correa:2012at} and Bethe-Salpeter resummation of the perturbative expansion \cite{Correa:2012nk}.  This result constitutes one of the few interpolating functions of $AdS_5/CFT_4$, i.e. functions that are exactly known at any value of the coupling constant and which therefore are of particular interest for probing the theory away from the perturbative reaches of the strong/weak coupling duality.\\ 

The idea in \cite{Correa:2012at} is to relate the cusp anomalous dimension  $\Gamma(\gl,\gamma,\gth)$ at $\gamma=0$ and $\gth\ll 1$ to the derivative of Wilson loops $\left<W(\gth=0)\right>$ that couples scalar fields to the equator $\gth=0$ of $S^2\in S^5$. Latitude loops in $\cN=4$ SYM were first investigated in perturbation theory in \cite{Drukker:2006ga} and were shown to be $\ov{4}-$BPS operators for $\gth\neq 0$. The amount of supersymmetry they preserve is enough for localization to apply \cite{Pestun:2009nn}. Their expectation value is equivalent at any order to that of the $\half$ BPS circular loop

$$
\left<\cW(\gl,\gth)\right> = \left<\cW(\gl',0)\right>
$$

provided the rescaling of the coupling constant $\gl'\to \gl\cos^2\gth$, as originally conjectured in \cite{Drukker:2006ga}. \\

In the context of the $AdS_4/CFT_3$ correspondence much less is known about the cusp anomalous dimension and the related function $h(\gl)$ \cite{Nishioka:2008gz,Gaiotto:2008cg,Grignani:2008is}. From the point of view of integrability, this function appears in the problem of determining the dispersion relation of a single magnon  \cite{Beisert:2005tm}

$$
E(p) = \sqrt{Q^2 +4\, h^2(\gl) \,\sin^2 \frac{p}{2}} - Q
$$

where $p$ is the magnon's momentum along the spin chain and $Q$ is its $R-$charge. It also appears in the TBA equations of \cite{Gromov:2008qe} that in turn determine the all-order spectrum of anomalous dimensions of gauge theory operators. Besides, the cusp anomalous dimension also relates the $AdS$ spin $S$ and scaling dimension $\Delta$ of twist-two operators 

$$
\Delta-S= \Gamma(\gl)\log S \quad {\rm for} \quad S \to \infty \quad {\rm and} \quad  \frac{\gamma^2}{2}\Gamma(\gl) = \lim_{\gamma\to\ii\infty} \Gamma(\gl,\gamma,0)
$$

and must be recovered by the appropriate semiclassical limit of the TBA system of equations. The original mismatch between the next-to-leading value attributed to $h(\gl)$ at $\gl\bb 1$ by integrability and the one obtained in the field-theoretic computation of \cite{Alday:2008ut} was solved in \cite{McLoughlin:2008he}. There the authors proposed to eliminate the dependence on the unphysical function $h(\gl)$ by re-expressing the $\gl$ dependence in terms of the physical $\Gamma(h(\gl))$ inside the TBA equations. \\

It was suggested in  \cite{Correa:2012at} that a comparison between the all-order expressions for the ABJM cusp anomalous dimension coming from integrability and a genuine gauge-theoretic computation should clarify the nature of the infamous function $h(\gl)$. In this paper we investigate the second half of that proposal.
In a first instance we construct, in Euclidean signature, the Wilson loop operator that couples gauge fields to a latitude at an angle $\phi$ of $S^2\in S^3$ and scalar fields to a latitude at an angle $\gth$ in $S^2\in \mathbb{C}P^3$ and study its BPS character for general values of $\phi,\gth$. We find that, admitting one identifies $\phi$ and $\gth$, it preserves 2 out of 12 superconformal charges and give explicit expressions for the Killing spinors parametrising them. We study such operator at weak coupling and note a striking similarity with  the four-dimensional result of \cite{Drukker:2006ga}. Namely, at the second order of perturbation theory we find    

$$ \left< W(\gl,\gth)\right> \sim 1+  \pi^2 \gl^2  \cos^2\gth - \frac{1}{6}\pi^2\gl^2 \qquad \gl\ll 1 $$

where the last term is the topological contribution of Pure Chern-Simons theory. Building on the analysis of perturbation theory we conjecture that the strong coupling v.e.v. of the Wilson loop should display the same rescaling of the 't~Hooft coupling that happens in the four dimensional theory.  We compute the expectation value at strong coupling by means of the classical type IIA superstring in $AdS_4\times \mathbb{C}P^3$ that ends on our contour located at the boundary of $AdS_4$ and on a latitute at an angle $\gth$ on a two-sphere fibrated over a maximal arch of $ \mathbb{C}P^3$. Relying on the localization result of \cite{Kapustin:2009kz} and the matrix model technology developed in \cite{Marino:2009jd,Marino:2011eh} we find  a striking agreement with this conjecture

$$ \left< W(\gl,\gth)\right> \sim \e^{\pi\cos\gth\sqrt{2\gl}}  $$

 This acts as a motivation for relating the bremsstrahlung function, defined through  $\Gamma(\gl,\gamma,\gth) \sim (\gamma^2-\gth^2) B(\gl) $ for $\gamma\sim\gth\sim 0$ to the derivative of the all-order expression for the $\half$ BPS Wilson loop computed in \cite{Klemm:2012ii}. In turn this gives a non-perturbative-in-$\gl$ expression for the function $B(\gl)$ in three dimensional ABJM theory

$$ B(\gl) = N\,c \left[ \frac{{\rm Ai'}(c(\gz-\xi))}{{\rm Ai}(c(\gz-\xi))} -\frac{{\rm Ai'}(c\gz)}{{\rm Ai}(c\gz)}  \right] - \pi\gl\cot \pi\gl+\ii\pi\gl+1  $$

being 

$$
c= \left( \frac{\pi^2}{2}k \right)^{\ov{3}} \qquad \gz= N-\frac{k}{24}-\frac{1}{3k} \qquad \xi= \frac{6}{3k}
$$

and Ai$(x)$ is the Airy function. The latter is a compact expression that encodes all $1/N$ corrections, and by writing it we tacitly assume that it should hold the large $N$ limit only. The such written bremsstrahlung function can be expanded at weak and strong coupling, this yields to

\be\nonumber
\begin{split}
& \gl\ll 1 \qquad B^{(1/2)}(\gl) = \pi^2 \gl^2 -\frac{2}{3}\pi^4 \gl^4 + \frac{17}{15}\pi^6 \gl^6 - \frac{5597}{2520}\pi^8 \gl^8 +\frac{481003}{45360}\pi^{10} \gl^{10} + \cO(\gl^{12}) \\
& \gl\bb 1 \qquad B^{(1/2)}(\gl) = \frac{\pi}{2} \sqrt{2\gl} + 1- \frac{2}{\pi} + \frac{\pi}{48\sqrt{2\gl}} + \cO(\gl^{-1})
\end{split}
\ee

We observe a striking similarity between the expansions above and the thought form of the function $h(\gl)$ both at weak and strong coupling, and we end with a comment on this relation. Conventions, additional computations and a brief synopsis of localization results for ABJM Wilson loops are in appendix.

%%%%%%%%%%%%%%%%%%%%%%%%%%%%%%%%%%%%%%%%%%%%%%%%%%%%%%%%%%%%%%%%%%%%%%%%%%%%%%%%%%%%%%%

\section{Deforming the scalar couplings}
\label{sec:weak}

The $\frac{1}{6}$ BPS bosonic Wilson loop of the $\cN=6$ ABJM theory  is known to couple the scalar fields to a semicircle inside $\mathbb{C}P^3$  \cite{Drukker:2008zx,Rey:2008bh}, in contrast with the four dimensional case in which the $\half$ BPS loop operator is coupled to a single point in $S^5$. There are no known purely bosonic loop operators that preserve more than one-sixth of the $\cN=6$ superconformal symmetry of ABJM theory, and one has to include the direct coupling of fermionic fields to the contour to obtain a $\half$ BPS observable \cite{Drukker:2009hy}. Thought the Wilson loop of  \cite{Drukker:2008zx,Rey:2008bh} is somewhat closer to the $\ov{4}$ operator of \cite{Drukker:2006ga} in $\cN=4$ SYM in four dimensions, where scalar fields are coupled to a latitude on a $S^2\in S^5$ at some angle $\gth$. We  are then interested in deformations of the operator of  \cite{Drukker:2008zx,Rey:2008bh}  that include the coupling of scalar fields to a generic arch in $\mathbb{C}P^3$ and still preserve (at least part of) its supersymmetry.

\subsection{BPS constraints on $S^3$}

The bosonic Wilson loop on $S^2\times \mathbb{R}$ in Euclidean signature reads

\be
W = \frac{1}{N} \Tr\cP \exp\int_\text{ \calligra l \normalfont }\left( \ii A_\mu\dot x^\mu + \frac{2\pi}{k} C_I M^I_{~J} \bar C^J \right)\dd\tau
\ee  

We use $ \mathbb{R}^3$ flat coordinates $x_1,\,x_2,\,x_3$ for the embedding of $S^2$, whose radius is set to one without any loss of generality and can be reintroduced by dimensional analysis. 
The loop operator is coupled to the contour 

\be
\label{x-contour}
x^\mu(\tau) = \{ \cos\phi\,\cos\tau,\, \cos\phi\,\sin\tau,\, \sin\phi \}
\ee

that parametrises a latitude \calligra l~ \normalfont on the two-sphere at an angle $\phi$ and with affine coordinate $\tau\in[0,2\pi)$. In an alternative description, the $S^2$ can be viewed as a maximal two-sphere in $S^3$; the two descriptions are equivalent and we will use either of them according to which is more convenient. The matrix $M^I_J$ determines the coupling of scalar fields to a contour in $\mathbb{C}P^3$ through the parametrization

\be
M^I_{~J} = \gd^I_J|\dot y| - 2\frac{\dot{\bar{y}}^I \dot y_J}{|\dot y|}
\ee

where $y_I,\bar{y}^I,\,I=1...4$ are projective coordinates in $\mathbb{C}P^3$. In the usual case where $M=\pm{\rm diag}(1,1,-1,-1)$, $y^I$ define a maximal semicircle. Supersymmetry requires 

\be
\label{x-and-y}
|\dot x|=|\dot y|
\ee

which fact we will discuss in more details in the following, hence it is convenient to rescale $M$ as

\be
\label{M-in-terms-of-y}
M^I_{~J}=|\dot y|M'^I_{~J} =|\dot y|\left( \gd^I_J - 2\frac{\dot{\bar{y}}^I \dot y_J}{|\dot y|^2} \right)
\ee

From now on we will remove the prime from $M'$ and refer to it as the matrix of scalar couplings. According to the supersymmetric transformation rules of bosonic fields (\ref{susytrans}), the variation of the Wilson loop relative to the Poicar\'e subgroup of the superconformal group reads

\be
\begin{split}
\label{eq:variationW}
\gd W =& \gd \left( \ii A_\mu \dot x^\mu + \frac{2\pi}{k} |\dot y| M^I_{~J}C_I \bar C^J \right)\\
=& -\frac{4\pi}{k}\dot x^\mu \left( \varepsilon_{IJ}^\ga (\gg_\mu)_\ga^\gb \bar\psi_\gb^I \bar C^J + 
C_J \psi_I^\ga (\gg_\mu)_\ga^\gb \bar\vartheta_\gb^{IJ}  \right)
 + \frac{4\pi}{k} |\dot y| \left( M^I_{~J}C_I \psi_L^\ga \bar\vartheta_\ga^{JL} + 
M^I_{~J} \varepsilon_{IL}^\ga \bar\psi_\ga^L \bar C^J \right)\\
=& \frac{4\pi}{k} \left\{ C_J \psi_I^\ga \left[ \dot x^\mu (\gg_\mu)_\ga^\gb \gd^{J}_{~L} + |\dot y| \gd_\ga^\gb
M^J_{~L} \right] \bar\vartheta_\gb^{LI} + \varepsilon_{LI}^\ga \left[ \dot x^\mu (\gg_\mu)_\ga^\gb \gd^L_{~J} + |\dot y| \gd_\ga^\gb M^L_{~J} \right] \bar\psi_\gb^I \bar C^J \right\}
\end{split}
\ee

Note that in the last line the antisymmetry of $\gth$ has been used. Most importantly, one should keep in mind that there is no reality condition on superconformal Killing spinors in Euclidean signature, henceforth $\varepsilon$ and $\bar\vartheta$ should be regarded as independent spinors. 
%{\color{red} \sc[Multipling by $\ii$ all the terms coming from the variation of $A_\mu$, that is equivalent to multipling by $\ii$ all the $\gg$'s, we recover the result of Trancanelli-Faraggi's notes in Lorentzian signature, whereas using the properties of $\gth$'s and relabelling $\gth \leftrightarrow \bar \gth$ one recovers Drukker-Plefka-Young.]}  
We expect that circular loops preserve, at most, Poincaré-conformal mixed supercharges, hence we parametrise Killing spinors as

\be
\bar\vartheta_\ga^{IJ} = \bar\gth_\ga^{IJ} + (x\cdot \gg \, \bar\eta)_\ga^{IJ} \qquad \varepsilon_\ga^{IJ} = \gep_\ga^{IJ} - (\gz\,x\cdot \gg)_\ga^{IJ}
\ee

where $\bar\gth, \,\gep$ parametrise super-Poicaré transformations $\gd_P=Q^\ga_{IJ}\bar\gth^{IJ}_\ga,\,\gep^\ga_{IJ}\bar Q^{IJ}_\ga$, and $\bar\eta,\,\gz$ parametrise conformal transformations $\gd_C = S^\ga_{IJ}\bar\eta^{IJ}_\ga,\,\gz^\ga_{IJ}\bar S^{IJ}_\ga$. We will solve separately for $\bar\vartheta$ and $\varepsilon$. In what follows we work out the solution for $\bar\vartheta$, but the procedure is the same in the two cases. 
Using the explicit form of the contour, the first term in parenthesis in the last line of (\ref{eq:variationW}), up to $C_P \psi_J^\ga $, becomes

\be
\begin{split}
\label{eq:variation-1}
\gd W_1 \sim & \,\,|\dot y|M^P_{~I}\bar\gth_\ga^{IJ}  -(\gg_1 \gg_2)_\ga^\gg \, \bar\eta_\gg^{IJ}\gd^P_{~I}\, \cos^2\phi\\
+& \left[ (\gg_1)_\ga^\gg \bar\eta_\gg^{IJ} M^P_{~I} |\dot y| + (\gg_2)_\ga^\gg \bar\gth_\gg^{IJ}\gd^P_{~I} \right]\cos\phi\,\cos\tau\\
+& \left[ (\gg_2)_\ga^\gg \bar\eta_\gg^{IJ} M^P_{~I} |\dot y| - (\gg_1)_\ga^\gg \bar\gth_\gg^{IJ}\gd^P_{~I} \right]\cos\phi\,\sin\tau\\
+& \left[ (\gg_2)_\ga^\gb \gd^P_{~I} \cos\phi\, \cos\tau -(\gg_1)_\ga^\gb \gd^P_{~I} \cos\phi\, \sin\tau +
\gd_\ga^\gb M^P_{~I}|\dot y|  \right](\gg_3)_\gb^\gg \bar\eta_\gg^{IJ} \,\sin\phi
\end{split}
\ee

Given the functional dependence on the affine parameter $\tau$ along the loop, there are different ways to gather the various bits and make (\ref{eq:variation-1}) vanish, but at a first glance it is not obvious which one is best to take into account the right amount of supersymmetry preserved. So, in a first instance, we ask that $\gd W_1=0$ is satisfied by equating each line of (\ref{eq:variation-1}) above to zero independently. Either multiplying from the left the second line by $\gg_2$ or the third line from the right by $\gg_1$ brings to the same condition between Poicar\'e and conformal parameters, namely that

\be
\label{eq:condition-1}
 \bar\gth_\ga^{PJ} = |\dot y|(\gg_1 \gg_2)_\ga^\gb M^P_{~I} \bar\eta_\gb^{IJ} = \ii |\dot y|(\gg_3)_\ga^\gb M^P_{~I} \bar\eta_\gb^{IJ} 
\ee

Moreover, whenever $\phi\neq 0$ the last line of (\ref{eq:variation-1}) must vanish independently. To this purpose it is convenient to project onto the eigenstates of definite chirality with respect to the tangent direction  to the loop $\bar\eta_\pm = P_\pm \bar\eta$, using the projectors $P_\pm=\half (\mathbb{I} \pm \frac{\dot x^\mu \gg_\mu}{|\dot x|})$. 
Expanding in components and availing on the condition (\ref{x-and-y}), it is easy to see when the last line of (\ref{eq:variation-1}) vanishes for any value of the latitude angle $\phi$, indeed the latter can be written as

\be
\label{eq:all-prop-to-sin-phi}
\left[ \left(\pm \mathbb{I}_\ga^\gb + (\dot X^\mu \gg_\mu)_\ga^\gb \right) \gd^P_{~I} + \gd_\ga^\gb\, \left( M^P_{^I} \mp  \gd^P_{~I} \right) \right](\gg_3)_\gb^\gg \bar\eta_\gg^{IJ}\, \sin\phi
\ee

where $X^\mu =\{\cos\tau,\, \sin\tau,\,0 \}$ is the circle with unitary radius. The first summand above then is simply proportional to

\be
\mp (P_\pm)_\ga^\gb \, (\gg_3)_\gb^\gg \bar\eta_\gg^{IJ}
\ee

Since the projectors mix the upper and lower components of $\bar\eta$, if we look for a solution of this kind, the former cannot be independent of each other, otherwise (\ref{eq:all-prop-to-sin-phi}) would lead to inconsistencies. Thus we parametrise $\bar\eta_2 = z \bar\eta_1$, in such a way that the eigenvalue equation for $\eta$ reads

\be
P_\pm \gg_3 \left( \begin{array}{c} 1 \\ z \end{array} \right) \bar\eta = \half \left( \begin{array}{c} 1\pm\ii\,z\,\e^{-\ii\tau} \\ \pm\ii\e^{\ii\tau} - z \end{array} \right) \bar\eta = \left( \begin{array}{c} 1 \\ -z \end{array} \right) \bar\eta
\ee

which is solved by $z=0,\infty,\pm\ii\e^{\ii\tau}$, and for some complex number $\eta$  that accounts for a suitable normalisation. The solution $z=0$, as well as $z=\infty$, gives a vanishing $\bar\eta=0$; the two more solutions respectively give eigenspinors with 0 and 1 eigenvalues. Hence the constraint above is solved by

\be
\label{eq:condition-2}
M^{I}_{~J} =  \gd^{I}_{~J} \qquad (\bar\eta^{IJ}_\pm)_\ga = \left(\begin{array}{c} 
1 \\ \e^{-\ii(\tau\pm\frac{\pi}{2})}
 \end{array}\right) \,n^{IJ}\bar\eta 
\ee

where the plus/minus signs hold respectively for positive/negative chirality of $\bar\eta^{IJ}_\ga$. Moreover the $R-$symmetric structure completely factors out and can be encoded in the tensor $n^{IJ}$. 
Substituting (\ref{eq:condition-1})  into the first line of $\gd W_1$ and asking that it too vanishes on its own we find

\be
\label{eq:condition-3}
|\dot y|^2 M^P_{~I}\,M^I_{~L}\, \bar\eta_\ga^{LJ} = \cos^2\phi\, \bar\eta^{PJ}
\ee

which luckily is compatible with (\ref{eq:condition-2}) and, not really surprisingly, is independent of the latitude angle at which \calligra l~ \normalfont lies. Notice however that the chiral decomposition also implies

\be
\bar\gth_{\pm}^{PJ} = \pm \ii |\dot x|M^P_{~I} \bar\eta_{\pm}^{IJ}
\ee

so $\bar\vartheta$ is also chiral, as a matter of fact considering the projector operator along the loop yields to

\be
\begin{split}
& \half \left(\mathbb{I} + \dot X\cdot \gg \right) \bar\theta_- = 0 \qquad \half \left(\mathbb{I} + \dot X\cdot \gg \right) \bar\eta_- = 0 \\
& \half \left(\mathbb{I} - \dot X\cdot \gg \right) \bar\theta_+ = 0 \qquad \half \left(\mathbb{I} - \dot X\cdot \gg \right) \bar\eta_+ = 0
\end{split}
\ee

or otherwise stated $\bar\gth_+,\bar\eta_+$ are chiral, whereas $\bar\gth_-,\bar\eta_-$ are anti-chiral.
 Alternatively, which is perhaps the most transparent way for counting Killing spinors, we can rearrange terms in (\ref{eq:variation-1}) and consider the new (equivalent) constraints on superconformal spinors

\be
\begin{split}
&\left[ (\gg_1)_\ga^\gg \bar\eta_\gg^{IJ} M^P_{~I} |\dot x| + (\gg_2)_\ga^\gg \bar\gth_\gg^{IJ}\gd^P_{~I} + (\gg_2)_\ga^\gb (\gg_3)_\gb^\gg \bar\eta_\gg^{IJ} \gd^P_{~I} \sin\phi\, \right]\cos\phi\,\cos\tau=0
\\
&\left[ (\gg_2)_\ga^\gg \bar\eta_\gg^{IJ} M^P_{~I} |\dot x| - (\gg_1)_\ga^\gg \bar\gth_\gg^{IJ}\gd^P_{~I} - (\gg_1)_\ga^\gb (\gg_3)_\gb^\gg \bar\eta_\gg^{IJ} \gd^P_{~I} \sin\phi\,\right]\cos\phi\,\sin\tau=0
\end{split}
\ee

Multiplying the first by $\gg_2$ from the left we obtain a deformation of (\ref{eq:condition-1}) by a term that vanishes on the equator of $S^2$

\be
\label{eq:cond-1-deformed}
\bar\gth_\ga^{PJ} = \ii (\gg_3)_\ga^\gb \left( |\dot x|\,M^P_{~I} + \ii \gd^P_{~I}\sin\phi \right) \bar\eta_\gb^{IJ}
\ee

Inserting it in what is left of the variation, namely

\be
\begin{split}
\gd W_1 \sim &|\dot x|M^P_{~I}\bar\gth_\ga^{IJ}  -(\gg_1 \gg_2)_\ga^\gg \, \bar\eta_\gg^{IJ}\gd^P_{~I}\, \cos^2\phi  +
 M^P_{~I}|\dot x| (\gg_3)_\ga^\gg \bar\eta_\gg^{IJ} \,\sin\phi\\
 = & |\dot x|^2 \,M^P_{~I}\,M^I_{~L}\, \bar\eta_\ga^{LJ}-  \bar\eta^{PJ} \cos^2\phi 
\end{split}
\ee

we obtain again (\ref{eq:condition-3}), but without any condition onto the plus and minus components of the superconformal spinors, opposed to the previous case. There is also a third way to rearrange the contributions to $\gd W_1$, which gives again the same result as above.
The second contribution to the last line of (\ref{eq:variationW}) and proportional to $\gth$ can be worked out along the same line. Note that there is no supersymmetry for $|\dot x|\neq |\dot y$ because in the last line of (\ref{eq:variation-1}) we cannot define any projector onto chiral states.

%%%%%%%%%%%%%%%%%%%%%%%%%%%%%%%%%%%%%%%%%%%%%%%%%%%%%%%%%%%%%%%%%%%%%%%%%%

\subsection{Determining the Killing spinors}

We would like to provide explicit solutions to the set of equations (\ref{eq:condition-1}-\ref{eq:cond-1-deformed}) constraining the scalar coupling and Killing spinors. We first notice that when the contour \calligra l \normalfont~ lies at the equator of the two-sphere, meaning that $\phi=0$, the two sets of constraints found above become equivalent, as the condition upon the two components of the superconformal spinors disappears. We then recover the set of constraints found in \cite{Drukker:2008zx}

\be
|\dot y|^2\, M^2=\mathbb{I} \qquad {\rm~and} \qquad |\dot y|^2 \, M^K_{~P}\,M^I_{~Q}\, \bar\eta_\ga^{QJ} \gep_{IJKL} =  -\bar\eta^{IJ}\gep_{IJPL}
\ee

These are easily solved by breaking the (still untouched) $SU(4)_R$ symmetry  to $SU(2)\times SU(2)$ and considering $2\times 2$ sub-blocks of $M$, namely by letting $I,J=1,2$ and $K,L=3,4$. This way it is easy to see that choice of $M=diag(1,1,-1,-1)$ preserves the spinors $\bar\vartheta^{12}$ and $\bar\vartheta^{34}$.  In this case, according to (\ref{eq:condition-1}), the conserved Poincaré-conformal spinors have the simple form

\be
\bar\vartheta^{12}_\ga = \left[(x\cdot\gg)_\ga^\gb +\ii(\gg_3)_\ga^\gb  \right] \bar\eta^{12}_\gb \qquad \bar\vartheta^{34}_\ga = \left[(x\cdot\gg)_\ga^\gb -\ii(\gg_3)_\ga^\gb  \right] \bar\eta^{34}_\gb
\ee

and $\bar\eta$ fulfills the requirement (\ref{eq:condition-2}). This holds for both chiralities, hence these are two out of twelve bi-spinors, six $\bar\gth$'s and six $\bar\eta$'s, and the operator is hence 1/6 BPS, as it is known.\\

It is not hard to generalise this to the more general case where $\phi\neq 0$. Let us focus on the simpler case where $M$ is still diagonal. For general values of the latitude angle there is, in addition to the principal constraints, a further equation for $\bar\eta$ and $M$ (\ref{eq:cond-1-deformed}) and the requirement that $\bar\vartheta$ is a Killing spinor.
To find a solution for the first set of constraints we choose $M^1_1=M^2_2=|\dot y|^{-1}\cos\phi =1$. Hence the constraints in (\ref{eq:condition-3}) and (\ref{eq:cond-1-deformed}) are fulfilled for some $\bar\eta^{12}$ such that 

\be
\bar\eta^{12}_\ga \sim \left(
\begin{array}{c}
\bar\eta \\
\e^{-\ii(\tau+\frac{\pi}{2})} \bar\eta
\end{array} \right) 
\ee

with the appropriate normalization factor being $1/(2\eta\bar\eta)$. The equation for $\vartheta$ is solved for $M^3_3=M^4_4= -1$ and $\eta_{34+} = \e^{\ii(\tau\pm\frac{\pi}{2})}\, \eta_{34-}$. Hence the loop operator coupled to a latitude at an angle $\phi$ and with the scalar couplings

\be
\label{eq:M1/6}
M={\rm diag} (1,1,-1,-1)
\ee

borrowed from the infinite Wilson line is annihilated by superconformal transformations parametrised by

\be
\label{eq:conserved-chiral}
\bar\vartheta^{12}_+ = \left[(x\cdot\gg) +\ii |\dot x| (\gg_3) \right] \bar\eta^{12}_+ =
\left[ |\dot x|(X\cdot\gg) +\ii \gg_3\,\e^{-\ii\phi} \right] \bar\eta^{12}_+
\ee

and analogously 

\be 
\bar\vartheta^{34}_- = \left[ |\dot x|(X\cdot\gg) -\ii \gg_3\,\e^{+\ii\phi} \right] \bar\eta^{34}_-
\ee

with $\bar\eta^{12},\, \bar\eta^{34}$ given above and with chirality defined with respect to the projectors $P_\pm = 1\pm\dot X\cdot\gg$. Notice that in this case, opposed to the $\phi=0$ case, there is one additional condition on Killing spinors that relates their two components. In other words the chiral decomposition and following identification of the two components up to a contour dependent phase factor, that in turn is needed to solve the additional $(\sin\phi)$-dependent  constraint, chops half of the degrees of freedom. This choice of parametrization of the superconformal transformations then allows the loop to be invariant under a linear combination of 2 out of the 24 generators of the full superconformal group. But the BPS character of this Wilson loop is actually larger. To this end, note that allowing for the deformation in (\ref{eq:cond-1-deformed}), the condition upon the plus and minus components of the superconformal spinors disappears and the latter become genuine two-component spinors. There is then a doubling of the supersymmetry preserved by the operator with scalar coupling (\ref{eq:M1/6}), that hence is 1/6 of the vacuum symmetry, and whose generators are parametrised by

\be
\label{eq:1/6generators}
\bar\vartheta^{12}_\ga = \left[(x\cdot\gg)_\ga^\gb -\ii (\gg_3)_\ga^\gb \e^{\ii\phi}  \right] \bar\eta^{12}_\gb \qquad \bar\vartheta^{34}_\ga = \left[(x\cdot\gg)_\ga^\gb +\ii (\gg_3)_\ga^\gb \e^{-\ii\phi}  \right] \bar\eta^{34}_\gb
\ee

We can explicitly check  that $\bar\vartheta^{12}_\ga,\,\bar\vartheta^{34}_\ga$ are genuine complex Killing spinors. Indeed on $S^3$ we can consider usual $SU(2)$ left-invariant vector fields $i^k_\mu$, in terms of which the Killing equation reads\footnote{The spin connection on $S^3$ is $\gw_{ij}=\gep_{ijk} i^k$ with normalisation $i^j\,i_k = \gd^j_k$.}

\be
\label{eq:killing}
\begin{split}
\nabla_\mu \bar\vartheta &= (\del_\mu +\ov{8}\gep_{ijk}[\gg^i,\gg^j] i^k_\mu) 
\left[(x\cdot\gg) \pm\ii (\gg_3) \e^{\pm\ii\phi}  \right] \bar\eta\\
&= \gg_\mu \left(\mathbb{I} + \frac{\ii}{2}(x\cdot\gg)\mp \half  \gg_3\e^{\pm\ii\phi} \right)\bar\eta
\end{split}
\ee  

for any constant bi-spinor $\bar\eta$ which is fixed only by normalisation.\\

%%%%%%%%%%%%%%%%%%%%%%%%%%%%%%%%%%%%%%%%%%%%%%%%%%%%%%%%%%%%%%%%%%%%%%%%%%%%

\subsection{Deforming to non-maximal arches in $\mathbb{C}P^3$}

Now we extend the previous results to operators that preserve locally an $SU(2)\times SU(2)$ subset of the $SU(4)_R$. In these settings, the most general solution to (\ref{eq:condition-3}) is given by

\be
\begin{split}
&(M^1_1)^2 + M^1_2 M^2_1 = 1\\
&M^1_1+M^2_2 =0
\end{split}
\ee  

This results from asking that $\bar\eta^{12}$ only is preserved by the upper block. Analogous equations hold for $\eta^{34}$ in the lower block. This corresponds to rotating the scalar fields $C_2,\,C_3$ into each other and then act with a local $SO(4)$ rotation 

\be
R = \left( \begin{array}{c|c}
\begin{array}{cc}
\cos(\gth_1/2) & \e^{\ii\gth_2}\sin(\gth_1/2)  \\  -\e^{-\ii\gth_2}\sin(\gth_1/2) & \cos(\gth_1/2)  
\end{array} & 0\\
\hline
0 & \begin{array}{cc}
\cos(\gth_2/2) & \e^{\ii\gth_2}\sin(\gth_2/2)  \\  -\e^{-\ii\gth_2}\sin(\gth_2/2) & \cos(\gth_2/2) \end{array} 
\end{array}\right)
\ee

Finally we get

\be
\label{eq:Mtheta}
M(\gth_1,\gth_2)= \left( \begin{array}{c|c}
\begin{array}{cc}
\cos\gth_1 & \e^{\ii\gth_2}\sin\gth_1  \\  \e^{-\ii\gth_2}\sin\gth_1 & -\cos\gth_1  
\end{array} & 0\\
\hline
0 & \begin{array}{cc}
\cos\gth_1 & \e^{\ii\gth_2}\sin\gth_1  \\  \e^{-\ii\gth_2}\sin\gth_1 & -\cos\gth_1 
\end{array} 
\end{array}\right)
\ee

Such deformation of the scalar coupling was introduced in \cite{Forini:2012bb,Griguolo:2012iq} to study the generalised cusp anomalous dimension of ABJ(M) theories. The key point now is to identify the phase $\gth_2$ with the affine parameter along the loop contour $\tau$. The matrix above can still be diagonalised to (\ref{eq:M1/6}) but the transformation involved is local in $\tau$ and will affect both the scalar kinematical term and the Yukawa couplings in the ABJM action. Henceforth the expectation value of the loop operator will differ from the one in \cite{Drukker:2008zx,Rey:2008bh}. According to (\ref{M-in-terms-of-y}), the path in $\mathbb{C}P^3$ reads in projective coordinates

\be
\label{contour-in-y}
\dot y^I = \left\{ -\sin\frac{\gth_1}{2}\,\e^{-\ii\frac{\tau}{2}},\,\cos\frac{\gth_1}{2}\,\e^{\ii\frac{\tau}{2}},\, \sin\frac{\gth_1}{2}\,\e^{\ii\frac{\tau}{2}},\,-\cos\frac{\gth_1}{2}\,\e^{-\ii\frac{\tau}{2}} \right\}
\ee 

This path corresponds to a circle lying at an angle $\gth_1$ on a $S^2$ fibrated over one of the maximal circuses of $\mathbb{C}P^3$. In the next section we will compute the v.e.v. of the Wilson loop at both weak coupling, through a direct field theory computation, and at strong coupling through the semiclassical dual stringy solution.

%%%%%%%%%%%%%%%%%%%%%%%%%%%%%%%%%%%%%%%%%%%%%%%%%%%%%%%%%%%%%%%%%%%%%%%%%%%%%

\section{The Wilson loop at weak and strong coupling}

\subsection{Two-loops at weak coupling}
\label{sec:two-loops-circles}

To make contact with previously known results in ABJM theory at loop order \cite{Drukker:2008zx,Rey:2008bh}, we compute the expectation value of the latitude loop in the decompactification limit. It is well established that, in Feynman gauge in $d=3-2\gep$ dimensions, all one-loop contributions vanish and at two loops there are only three relevant diagrams -- the one-loop corrected Chern-Simons propagator, the double scalar exchange and the Chern-Simons three vertex. These can be categorized into the topological contribution, coming from CS fields only, and the matter contribution relative to the exchange of matter fields at any loop order, thus including corrections to the topological sector as well. In our case the matter sector contribution at 2-loop order is 

\be
\label{eq:Wmatter}
\begin{split}
\left< W_{\rm matter}^{(2)}\right>=& \int\dot x_1^\mu \dot x_2^\nu \left< A_\mu(\tau_1) A_\nu(\tau_2) \right> + \left( \frac{2\pi}{k} \right)^2 |\dot x_1| |\dot x_2| \left< [M^I_J C_I\bar C^J](\tau_1)[M^K_L C_K\bar C^L](\tau_2) \right>\\
=& \int\left(\frac{2\pi}{k} \right)^2 \left(\frac{N}{2\pi}\right)^2 \left\{ \frac{\dot x_1\cdot \dot x_2 -\frac{1}{4} \Tr M(\tau_1)M(\tau_2)}{(x(\tau_1)-x(\tau_2))^2}  \right\}
\end{split}
\ee 

where the shorthand $[M^I_J C_I\bar C^J](\tau_1)$ is understood as the value of the composite operator $MC\bar C$ in $x(\tau_1)$, being $\tau$'s the affine parameters on the latitude. Using the matrix of scalar couplings (\ref{eq:Mtheta}) and idetifing the angular parameter $\gth_2$ with the affine coordinate we get

\be
= \int_0^{2\pi}d\tau_1\,d\tau_2\,\left(\frac{N}{k} \right)^2 \frac{\cos(\tau_1-\tau_2) - \cos^2\gth_1-\sin^2\gth_1\,\cos(\tau_1-\tau_2)}{2[1-\cos(\tau_1-\tau_2)]} = \pi^2 \left(\frac{N}{k} \right)^2  \cos^2\gth_1
\ee

Comparing the result above with its counterpart in \cite{Drukker:2006ga}, we note that the two differ by a rescaling of the 't~Hooft coupling constant $\gl\to\cos^2\gth_1\gl$ which is strongly reminiscent of what happens in the four dimensional $\cN=4$ SYM theory \cite{Drukker:2006zk,Drukker:2007dw,Drukker:2008zx}. This fact originally led to conjecture that such rescaling of the coupling constant would persist at strong coupling, which  was indeed later proved in \cite{Pestun:2009nn} using localization. The present case is somewhat different though. To get the full 2-loop contribution we must add the topological contribution of the CS-three vertex $\left< W_{\rm top}^{(2)}\right> = - \frac{1}{6}\pi^2\gl^2$ to the result above. The vev of the Wilson loop then becomes

\be
\label{2-loop-gauge-theory}
\left< W(\gl,\gth)\right>= 1+  \pi^2 \gl^2  \cos^2\gth_1 - \frac{1}{6}\pi^2\gl^2
\ee

Building on this expression we conclude that an effective rescaling of the 't~Hooft coupling at strong coupling, eventually, cannot follow the simple pattern of the four dimensional case. It is indeed the fact that the $\cN=4$ matrix model is captured by the sum of infinite ladder diagrams that determines the exponentiation of the rescaling. As it was shown in \cite{Rey:2008bh}, the resummation of the composite gauge-scalar two-loop propagator would predict a strong coupling behaviour of the form $\exp(\cos\gth_1\,\gl)$ which is incompatible with the leading asymptotics of the dual string solution  $\left< W\right> \sim \exp\sqrt{\gl}$ \cite{Drukker:2006ga}.\\

On the other hand we are led to conjecture that a slight modification of that argument should hold in the present case. To this end we consider ABJM theory in light cone gauge $A_-=A_0-A_1=0$. Due to the anti-symmetry of the CS three-vertex, pure CS theory in this gauge is a theory of free propagators \cite{Frohlich:1989gr,AlvarezGaume:1989wk}. The one-loop corrected CS propagator was computed in \cite{Marmiroli:2012ny} and differs only by a total derivative from the Feynman gauge one. It appears natural to separate the contribution of the matter and topological sectors 

\be
\label{non-pert-guess}
\left< W(\gl,\gth)\right> = \e^{t(\gl) + m(\gl,\gth)}
\ee

In such way the perturbative expansion can be expressed diagrammatically in a very convenient and intuitive (at least at the first few orders) way  

\be
\begin{split}
\left< W(\gl,\gth)\right> =& 1 + \gl \left( \raisebox{-.03\textwidth}{\includegraphics[height=.08\textwidth]{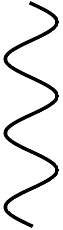}} + \raisebox{-.03\textwidth}{\includegraphics[height=.08\textwidth]{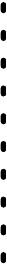}}\,\, \right)+ \half \gl^2 \left(  \raisebox{-.03\textwidth}{\includegraphics[height=.08\textwidth]{gauge.pdf}} + \raisebox{-.03\textwidth}{\includegraphics[height=.08\textwidth]{scalar.pdf}}\,\, \right)^2 + \dots\\
 =& 1 + \underbrace{%
\gl \,\, \raisebox{-.03\textwidth}{\includegraphics[height=.08\textwidth]{gauge.pdf}} \,\, + \half \gl^2   \,\, \raisebox{-.03\textwidth}{\includegraphics[height=.08\textwidth]{gauge.pdf}}  \raisebox{-.03\textwidth}{\includegraphics[height=.08\textwidth]{gauge.pdf}} \,\, + \,\, \dots
}_{t(\gl)}
 + \underbrace{%
 \gl \,\, \raisebox{-.03\textwidth}{\includegraphics[height=.08\textwidth]{scalar.pdf}} \,\, + \half \gl^2 \left( \,\, \raisebox{-.03\textwidth}{\includegraphics[height=.08\textwidth]{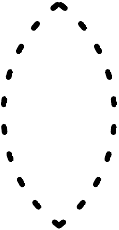}} \,\,+\,\,  \raisebox{-.03\textwidth}{\includegraphics[height=.08\textwidth]{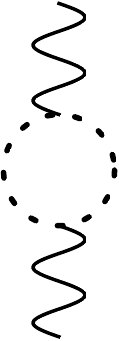}} \,\, \right) + \dots
}_{m(\gl,\gth)}
\end{split}
\ee

Therefore the two functions $t(\gl)$ and $m(\gl,\gth)$ are given respectively by the sum of CS tree level propagators in the first case and matter propagators, corrected gauge and (effective) Yukawa couplings in the second case. Notice that fermions will appear at higher orders even though they do not couple directly to the contour \footnote{This suggests that a reasoning along this line should be better defined in the superloop case.}. The factorisation in (\ref{non-pert-guess}) is somewhat similar to the proposal of \cite{Rey:2008bh}, but in the present case the matter contribution does not only include combined gauge-scalar two-loop propagators, but also interactions. At strong coupling, $m(\gl,\gth)$ is dominated by the exchange of the four-dimensional-like contributions in (\ref{eq:Wmatter}). The contribution of $SU(N)$ pure Chern-Simons theory is well known 

\be
\label{Wtopological}
\left< W_{\rm top}(\gl)\right> = \e^{-\pi\ii\gl}\frac{\sin \pi\gl}{N \sin\pi\gl/N}
\ee

therefore at large $\gl$ and in the planar approximation the Wilson loop should be dominated by 

\be
\log\left< W\right> \sim m(\gl,\gth_1)\sim f(\cos\gth_1)\sqrt{2\gl}
\ee

with some function $f$. From the large $\gl$ asymptotics we deduce that the contribution of composite gauge-scalar propagators does not dominate in that region. So we expect that interaction diagrams arising from Yukawa couplings will dominate at strong coupling, and these typically have more insertions of the loop coupling $M^I_J$ than of the coupling constant, so the expected large $\gl$ rescaling has the form 

\be
\label{predicted-rescaling}
f^2(\cos\gth_1)<\cos\gth_1
\ee

We will see from the dual string computation of next section that this is indeed the case.

%%%%%%%%%%%%%%%%%%%%%%%%%%%%%%%%%%%%%%%%%%%%%%%%%%%%%%%%%%%%%%%%%%%%%%%%%%%%

\subsection{Fundamental string solution}
\label{fundamental-string}

To analyse the behaviour of the loop operator in the large 't~Hooft coupling region we exploit the duality between the superconformal $\cN=6$ ABJM theory in three dimensions and type IIA superstring theory on $AdS_4\times \mathbb{C}P^3$.  Namely, we compute the classical area of the open string that ends on the contour $x^\mu(\tau)$ in (\ref{x-contour}) on the boundary of $AdS_4$ and on $y^I(\tau)$ in (\ref{contour-in-y}) inside $\mathbb{C}P^3$. The supergravity background includes the metric, the dilaton field, 2-form and 4-form field strengths, but at classical level we only need the metric and the relation between the $AdS$ radius and the gauge theory 't~Hooft coupling in units of $\ga'$

\be
\frac{R^3}{4k} = \pi\sqrt{2\gl}
\ee

The radius of  $\mathbb{C}P^3$ is twice the radius of $AdS_4$, so we use the convention 

\be
\dd s^2 = \frac{R^3}{4k}\left(\dd s^2_{AdS_4} + 4\dd s^2_{\mathbb{C}P^3}  \right)
\ee

The $AdS_4$ metric in global coordinates reads

\be
\dd s^2_{AdS_4} = -\cosh^2\rho \,\dd t^2 +\dd\rho^2 + \sinh^2\rho\,(\dd\phi^2+\sin^2\phi\,\dd\psi^2)
\ee

Also, it is convenient to adopt $\mathbb{C}^4$ projective coordinates to parametrise  $\mathbb{C}P^3$

\be
\begin{split}
&z_1= \cos\frac{\ga}{2}\cos\frac{\gb_1}{2}\e^{\ii\frac{\varphi_1}{2}+\ii\frac{\chi}{4}}\\
&z_2= \cos\frac{\ga}{2}\sin\frac{\gb_1}{2}\e^{-\ii\frac{\varphi_1}{2}+\ii\frac{\chi}{4}}\\
&z_3= \sin\frac{\ga}{2}\cos\frac{\gb_2}{2}\e^{\ii\frac{\varphi_2}{2}-\ii\frac{\chi}{4}}\\
&z_4= \sin\frac{\ga}{2}\sin\frac{\gb_2}{2}\e^{-\ii\frac{\varphi_2}{2}-\ii\frac{\chi}{4}}\\
\end{split}
\ee

The ranges of the angles are $0\leq\ga,\gb_{1,2}\leq\pi,\,0\leq\varphi_{1,2}\leq 2\pi,\,0\leq\chi\leq 4\pi$. These lead to the usual Fubini-Study metric

\begin{multline}
4\dd s^2_{\mathbb{C}P^3} = \dd\ga^2 + \cos^2\frac{\ga}{2}\left( \dd\gb_1^2 + \sin^2\gb_1\,\dd\varphi_1^2 \right) +\sin^2\frac{\ga}{2}\left( \dd\gb_2^2 + \sin^2\gb_2\,\dd\varphi_2^2 \right)\\ + \sin^2\frac{\ga}{2}\cos^2\frac{\ga}{2}\left( \dd \chi + \cos\gb_1\,\dd\varphi_1 - \cos\gb_2\,\varphi_2 \right)^2
\end{multline}

Let us introduce string worldsheet coordinates  $\tau,\gs$ parametrising respectively the compact and non-compact directions. We use the following ansatz in $AdS_4$ for the classical solution

\be
\label{string-ansatz}
\rho=\rho(\gs) \qquad  \psi(\tau)=\tau \qquad \phi=\phi(\gs),\,\phi(\gs_0)=\frac{\pi}{2}-\phi_0 \qquad t={\rm const.}
\ee

and in $\mathbb{C}P^3$ 

\be
\gb_1=\gb_1(\gs),\,\gb_1(\gs_0)=\gth_0 \qquad \varphi_1(\tau) = \tau \qquad  \ga,\,\varphi_2,\,\gb_2,\,\chi =0
\ee

Here we denote with $\phi_0,\,\gth_0$ the two angles defining the couplings of the Wilson loop operator to distinguish them from the ten dimensional coordinates.  The area is regularised introducing the cutoff $\gs_{min}$ that corresponds to cutting the hyperbolic radial coordinate at some $\rho_{\rm max}$. Note that with this choice the $\dot y^I(\tau)$ in (\ref{contour-in-y}) become simply 

\be
\dot y^I=\{ -z_2,\,z_1,\,z_2,\,-z_1\}
\ee

We can now write down the action 

\be
S= \frac{1}{2\pi} \frac{R^3}{4k} \int\dd\tau\,\dd\gs\,\left(\rho'^2 + (\phi'^2+\sin^2\phi)\sinh^2\rho + \gb_1'^2 + \sin^2\gb_1  \right)
\ee

and the Virasoro constraints

\be
\rho'^2 - (\sin^2\phi - \phi'^2)\sinh^2\rho=0 \qquad \gb_1'^2 - \sin^2\gb_1 =0
\ee

from which we read the equations of motion

\be
\begin{split}
&\rho'' = \sinh\rho\,\cosh\rho\,(\phi'^2+\sin^2\phi)\\
&\phi'' = \sin\phi\,\cos\phi\\
&\ga'' = -2\sin\frac{\ga}{2}\,\cos\frac{\ga}{2}
\end{split}
\ee

The first integrals of motion are easy to evaluate, we first integrate to find $\phi(\gs)$

\be
\sin\phi(\gs)=\frac{1}{\cosh(\gs_0\pm\gs)} \qquad \cos\phi_0 = \frac{1}{\cosh\gs_0}
\ee

where the integration constant $\gs_0$ is fixed by boundary conditions. Then inserting the latter into the integral for $\rho(\gs)$ we have

\be
\sinh\,\rho(\gs)=\frac{1}{\sin[\sqrt{2}(\gs_0\pm\gs)]}
\ee

with the boundary condition $\sinh\rho_{max} = \sin^{-1}\gs_{min}$. Analogously for the angular variable in $\mathbb{C}P^3$ 

\be
\sin\gb_1(\gs)=\frac{1}{\cosh(\gs_0\pm\gs)} \qquad \cos\gth_0=\tanh\gs_0
\ee

Finally, identifying $ 2\phi_0=\gth_0$ as in the gauge theory, substituting into the action and integrating with respect to the worldsheet variables we find

\be
\begin{split}
S_{cl} &= \frac{1}{2\pi} \frac{R^3}{4k} \int_0^{2\pi}\int_{\gs_{min}}^{\gs_{max}}\dd\tau\,\dd\gs\, \left[ \frac{1}{\sin^2[\sqrt{2}(\gs_0\pm\gs)]}+1 \right]\frac{1}{\cosh^2(\gs_0\pm\gs)}\\
&\sim\pi\sqrt{2\gl} \left[\cosh\rho_{max}\pm 1  \right]\cos\gth_0 + {\rm subleading}
\end{split}
\ee

The first, divergent contribution cancels against a boundary term as usual, whereas the finite part determines the large $\gl$ behaviour of the Wilson loop

\be
\label{loop-semiclassical-string}
\left< W \right> \sim \e^{\pm\pi\cos\gth_0\sqrt{2\gl}}
\ee

Since evidently  $\cos^2\gth_0 < \cos\gth$ in the whole interval $0<\gth_0<\pi$, this result is in agreement with the prediction made in (\ref{predicted-rescaling}) that, due to interacting diagrams, the strong coupling rescaling on $\gl$ is harder then what observed at weak coupling.

%%%%%%%%%%%%%%%%%%%%%%%%%%%%%%%%%%%%%%%%%%%%%%%%%%%%%%%%%%%%%%%%%%%%%%%%%%%%%

\section{An all-order expression for the Wilson loop at $\gth\sim\phi\sim 0$}

We are interested in computing the quantum average

\be
\label{eq:loop-path-integral}
\left<W(\gth) \right> = \left< \Tr\cP\,\e^{\oint \{A_\mu \dot x^\mu+\frac{2\pi}{k}M(\gth,\tau)C\bar C\}\dd\tau} \right>
\ee

with scalar couplings determined by the matrix (\ref{eq:Mtheta}). We propose that the latter equals the expectation value of the Wilson loop with winding number $n=\cos\gth$ in the matrix model obtained by localising the theory on $S^3$ \cite{Kapustin:2009kz}

\be
\label{eq:loop-winding}
\left< W(\gth)\right> = \int\dd\mu \rho(\mu)\e^{\cos\gth\,\mu}
\ee

being $\rho(\mu)$ the density of eigenvalues. This proposal is motivated by the fact that it is possible to put the scalar couplings given by $M(\gth)$ in diagonal form acting with an affine transformations on scalar fields. It can be seen that this transformation does not affect nor the kinetic term of scalars, neither the equations of motion of the auxiliary scalar field of the vector multiplet used in  \cite{Kapustin:2009kz} to localize the Wilson loop (see appendix). On this side of course a more rigorous computation would be needed, in particular concerning the fate of the one-loop determinants under this transformation. 
Equation (\ref{eq:loop-winding}) is a complex integral over a contour embracing the two cuts located on the intervals $\cC_1=[-A,A]$ and $\cC_2=[-B+\ii\pi,B+\ii\pi]$. The endpoints of the two cuts are defined through \cite{Marino:2009jd}

\be
a=\log A \qquad b=\log B \qquad  \ga=a+\ov{a} \qquad \gb=b+\ov{b}
\ee

and the mirror map

\be
\begin{split}
a(k) &=\half\left( 2+\ii \kappa+\sqrt{\kappa(4\ii-\kappa)} \right)\\
b(k) &=\half\left( 2-\ii \kappa+\sqrt{-\kappa(4\ii+\kappa)} \right)
\end{split}
\ee

The mirror map also determines the 't~Hooft coupling in terms of its mirror $\kappa$

\be
\label{mirror-map}
\gl(\kappa)= \frac{\kappa}{8\pi} ~_3F_2\left(\half,\half,\half;1,\frac{3}{2};-\frac{\kappa^2}{16}  \right)
\ee

In what follows we will check expression (\ref{eq:loop-winding-n}) at weak coupling against the gauge theory computation of Section \ref{sec:two-loops-circles} and at strong coupling against the classical string solution derived in Section \ref{fundamental-string}.

%%%%%%%%%%%%%%%%%%%%%%%%%%%%%%%%%%%%%%%%%%%%%%%%%%%%%%%%%%%%%%%%%%%%%%%%%%%%%

\subsection{Weak coupling expansion}
\label{weak-coupling-expansion}

 The density of eigenvalues for the ABJ matrix model was computed in \cite{Marino:2009jd} availing on the relation between the analytical continuation of the matrix model derived in \cite{Kapustin:2009kz} and the Chern-Simons matrix model on the lens space $L(2,1)$, see Appendix \ref{sec:CS-localization}. Changing integration variable to $x=\e^\mu$, it is then possible to write (\ref{eq:loop-winding-n}) explicitly as 

\be
\label{thetaloop}
\left< W(\gth)\right> = \frac{1}{2\ii\pi^2\gl(\kappa)}\int_{1/a}^a \dd x\,x^{\cos\gth-1} \tan^{-1}\left[\frac{\sqrt{\ga x-1-x^2}}{\sqrt{\gb x+1+x^2}}  \right]
\ee

For $\gth=0$, the derivative of this integral w.r.t. the mirror coupling $\kappa$ can be done exactly in terms of elliptic integrals \cite{Marino:2009jd}

\be
\label{mm-weak-master-integral}
\begin{split}
&\frac{\d}{\d(\ii\kappa)}\int_{1/a}^a \dd x\, \tan^{-1}\left[\frac{\sqrt{\ga x-1-x^2}}{\sqrt{\gb x+1+x^2}}  \right] = \half \int_{1/a}^a \frac{x\,\dd x}{\sqrt{(\ga x-1-x^2)(\gb x+1+x^2)}}\\
=& -\frac{1}{\sqrt{ab}(1+ab)}\left( a \mathbb{K}(m) -(a+b)\Pi(n|m) \right)
\end{split}
\ee

where the modulus and parameter are

\be
m^2 = \frac{(a^2-1)(b^2-1)}{(1+ab)^2} \qquad n=\frac{b}{a}\frac{a^2-1}{1+ab}
\ee

In the present case we cannot integrate the r.h.s. of the first line of (\ref{mm-weak-master-integral}) exactly, but we can expand at small $\gth$ and get

\be
\label{master-expanded}
x^{\cos\gth} = x - \frac{\gth^2}{2}x\,\log x + \cO(\gth^4) 
\ee

The first term generates the same integral as above, while the second term can be first expanded in small $\kappa$ then integrated in $x$ and again integrated in $\kappa$ to get the contribution to (\ref{thetaloop})

\be
\label{theta-term-weak-expansion}
\frac{\gth^2}{8 \pi^2\ii\gl} \int\dd\kappa\,\int_{1/a}^a \frac{x\,\log x \,\dd x}{\sqrt{(\ga x-1-x^2)(\gb x+1+x^2)}} = \ii\pi  \theta^2 \lambda +\frac{1}{2} \ii \pi^3 \theta^2 \lambda^3 -\frac{7}{36} \ii \pi^5 \theta^2 \lambda^5 + \cO(\gl^7)
\ee

Here we have used the inverse of the  mirror map (\ref{mirror-map}) to express the Wilson loop as a function of the 't~Hooft coupling.
There is no formal obstruction to carry this procedure forward to arbitrary loop order. The contribution of (\ref{theta-term-weak-expansion}) must be added to the term coming from the zero-framing integral (\ref{mm-weak-master-integral}). Consider that the framing factor of a loop with winding number $n$ is actually $n^2$, so the expectation value of (\ref{eq:loop-path-integral}) for $\gl\ll 1$ reads

\be
\label{theta-loop-weak-from-mm}
\begin{split}
\left< W(\gl,\gth) \right> &= \e^{\ii\pi(1-\gth^2)\gl} \left( 1+ (1-\gth^2)\pi^2\gl^2 -\ov{6}\pi^2\gl^2 - \ii\frac{5}{9} \left( 1-\frac{3}{2}\gth^2 \right)\pi^3\gl^3 +\frac{\ii}{18}\pi^3\gl^3 \right.\\
&\left.-\frac{1}{12}(1-2\gth^2)\pi^4\gl^4-\frac{19}{120}\pi^4\gl^4-\ii\frac{29}{450}\left(1-\frac{5}{2}\gth^2  \right)\pi^5\gl^5 - \ii\frac{133}{900}\pi^5\gl^5 +\cO(\gl^6) \right)
\end{split}
\ee

where we have gathered powers of $\cos^n\gth \sim 1-\frac{n\gth^2}{2}$ at order $\gl^n$. We note that at up to two-loop order this result is in precise agreement with the gauge theoretic computation (\ref{2-loop-gauge-theory}).  Moreover, one can separate in (\ref{theta-loop-weak-from-mm}) above the pure Chern-Simons contribution and the matter contribution, or otherwise stated, the perturbative expansions of the functions $t(\gl)$ and $m(\gl,\gth)$ in (\ref{non-pert-guess}) at zero framing

\be
\begin{split}
\left< W(\gl,\gth) \right> &= \e^{\ii\pi(1-\gth^2)\gl} \exp\left( (1-\gth^2)\pi^2\gl^2  - \ii\frac{1}{2} \left( 1-\frac{5}{6}\gth^2 \right)\pi^3\gl^3 -\frac{1}{12}(7-12\gth^2)\pi^4\gl^4 \right)\\
& \times \left(1-\ov{6}\pi^2\gl^2 + \ov{120}\pi^4\gl^4 \right) +\cO(\gl^5)
\end{split}
\ee

The second factor is evidently the perturbative expansion of pure Chern-Simons theory at framing zero. The expression above is in agreement with the factorisation property outlined in (\ref{non-pert-guess}), though one would need higher order results from the genuine gauge theory to actually prove exponentiation of the matter contribution. We would like to stress that indeed exponentiation is not needed in our current treatment, but it would certainly be interesting to investigate this point further.

%%%%%%%%%%%%%%%%%%%%%%%%%%%%%%%%%%%%%%%%%%%%%%%%%%%%%%%%%%%%%%%%%%%%%%%%%%%%%%

\subsection{Strong Coupling expansion}
\label{strong-coupling-expansion}

The expectation value of a Wilson loop with arbitrary winding number $n$ was computed in \cite{Klemm:2012ii} using the Fermi gas approach to the ABJM matrix model \cite{Marino:2011eh} and can be written in a compact form in terms of Airy functions. In the strongly coupled region the  $\frac{1}{2}$ BPS operator admits the expansion (at genus zero)

\be
\left< W^{(1/6)}_n \right> \sim \frac{1}{n} \e^{\pi n\,\sqrt{2\gl}} + \cO(\gl^0) 
\ee

 On the other hand it is straightforward to extract the large $\gl$ asymptotics directly from the integral (\ref{thetaloop}) after having set the two 't~Hooft coupling equal to each other.
For $\gl\to\infty$ the mirror map reads

\be
\gl(k) = \frac{1}{2\pi^2}\log^2 k +\frac{1}{24} +\cO(K^{-1})
\ee

then one can expand the integral for $k\to\infty$ and integrate. At the first subleading order in $\gl$ the result is 

\begin{multline}
\left< W(\gth)\right> = \frac{c_1}{\cos\gth} \e^{\pi\cos\gth\,\sqrt{2\gl}} \left[1+ (\cos\gth-1)\e^{-\pi\sqrt{2\gl}}  \right] \\
- \frac{c_2 \left(1+\cos\gth\e^{-\pi\sqrt{2\gl}}\right)}{\cos\gth(\cos\gth+1)}\, \e^{\pi(\cos\gth-1)\sqrt{2\gl}}
-\frac{c_1(\cos\gth-1)}{\cos\gth}\e^{-\pi\cos\gth\sqrt{2\gl}}
\end{multline}

being 

$$c_1=\coth(1) \qquad  c_2 = \frac{\e^4(\cos\gth+1)-8\e^2(2\cos\gth+3)-(\cos\gth+1)}{(\e^2-1)^2}$$

The asymptotic behaviour at $\gl\bb 1$ of the loop operator with winding number $\cos\gth$ reproduces quite well the semiclassical string computation (\ref{loop-semiclassical-string}). Also, it is in agreement with an analytic continuation of the result of \cite{Marino:2011eh} at non-integer winding number. This supports our conjecture that the expectation value of Wilson loop (\ref{eq:loop-path-integral}) at any value of the coupling constant $\gl$ is given, at least at small values of the parameter $\gth$, by the analytically continued result of \cite{Marino:2011eh} at winding number $\cos\gth$ and genus zero.

%%%%%%%%%%%%%%%%%%%%%%%%%%%%%%%%%%%%%%%%%%%%%%%%%%%%%%%%%%%%%%%%%%%%%%%%%%%%%%

\section{$\Gamma_{\rm cusp}(\gl,\phi,\gth)$ and the function $h(\gl)$}
\label{gamma-cusp}

It was recently proposed in \cite{Correa:2012at} to extract information about the all-order in $\gl$, small $\phi$ and $\gth$, cusp anomalous dimension $\Gg(\gl,\phi,\gth)$ in four dimensional $\cN=4$ SYM by relating the latter to the logarithmic derivative of the Wilson loop. In this section we proceed along the same line of \cite{Correa:2012at}. In a conformal theory, the Euclidean cusp anomalous dimension is related to the static potential between a couple of $W$-boson and anti-boson \cite{Drukker:2011za}. Because there are no quarks in this theory, {\it i.e.} no fields in the fundamental representation of the gauge group, $W$-bosons are obtained by higgsing the gauge group, which in turn is done by detaching a D-brane and taking it very far away from the other. Massive $W$'s that emerge from this procedure then reproduce the trajectories of quarks and anti-quarks, from which reason the potential can be thought of as the supersymmetric analog of the quark-anti-quark potential. In the case of ABJ(M) theory, $W$-bosons are $\half$ BPS particles obtained by suitably Higgsing $U(N+1)\times U(M+1)\to U(N)\times U(M)$ \cite{Lee:2010hk}. Note that although they  naturally couple to the superconnection of \cite{Drukker:2009hy}, localization establishes a remarkably simple relation between the $\half$ and $\ov{6}$ BPS operators (\ref{eq:relation1/2-1/6}).

\subsection{Extracting the cusp anomalous dimension}

The generalised Euclidean cusp anomalous dimension at $\phi\sim 0$ can be seen as coming from a conformal mapping of the $W\Wbar$ static potential 

\be
\Gamma(\gl,\gth,\phi=0) = \log\left< W_{W\Wbar}(\gl,\gth)\right>
\ee

where

\be
\left< W_{W\Wbar}(\gl,\gth) \right> = \left< \e^{\int_{-\infty}^\infty \ii A_\mu \dd x^\mu +\frac{2\pi}{k} M(0)C\bar C |\dd x| + \int_{\infty}^{-\infty} \ii A_\mu \dd x^\mu +\frac{2\pi}{k} M(\gth)C\bar C |\dd x| }  \right>
\ee

accounts for the interaction of a boson-anti-boson pair sitting at opposite sites in $S^2\times \mathbb{R}$ and with an internal angle differing by $\gth$. Here, the scalar coupling $M(\gth)$ are the ones obtained in (\ref{eq:Mtheta}). At first non-trivial order in $\gth$ we have

\be
\label{WWbar-pot}
\begin{split}
\left< W_{W\Wbar}(\gl,\gth) \right> =& 1+ \half \left( \frac{2\pi}{k} \right)^2 \int\int  \dd s\,\dd t\, \Tr[M(0)-M(\gth)]^2
\left< [C\bar C](s)[C\bar C](t) \right>\\
=& 1+ 2\gth^2 \left( \frac{2\pi}{k} \right)^2 \frac{N^2 \Gamma(1/2-\gep)^2}{\left( 4\pi^{\frac{3}{2}-\gep} \right)^2}  \int\int  \dd s\,\dd t\, \frac{1}{(s-t)^{2-4\gep}} 
\end{split}
\ee

where we have used the dimensionally regularised scalar propagator in $d=3-2\gep$ dimensions. Cutting of the integral at very large length $\pm L/2$, (\ref{WWbar-pot}) above becomes

\be
\left< W_{W\Wbar}(\gl,\gth) \right> = 1+ \frac{\gth^2}{2} \frac{N^2}{k^2} \frac{L^{4\gep}}{4\gep} +\cO(\gep)
\ee

On the other hand it was observed that the three dimensional cusp anomalous dimension is a BPS operator for $|\phi|=|\gth|$ at both strong \cite{Forini:2012bb} and weak coupling \cite{Griguolo:2012iq}, resembling in this its four-dimensional counterpart \footnote{Interestingly enough it was noted there that both the purely bosonic and the "super" operators are BPS for  $|\phi|=|\gth|$.}. From the BPS character of $\Gg(\gl,\phi,\gth)$  it follows that at $\phi\sim\gth\sim 0$ the cusp anomalous dimension itself must satisfy

\be
\Gg(\gl,\phi,\gth)= (\gth^2-\phi^2) B(\gl)
\ee

in this case being 

\be
B(\gl) = \half \gl^2 + \cO(\gl^4)
\ee

The function $B$  is related to the bremsstrahlung of soft gauge fields from heavy $W$ bosons undergoing a sudden change in direction by an angle $\phi$. This function was successfully determined in \cite{Correa:2012at} at all orders in the 't~Hooft coupling constant and matched with computations from integrability techniques \cite{Correa:2012hh} and Bethe-Salpeter resummation of the perturbative series \cite{Correa:2012nk}. Now, let us consider the small $\gth$ expansion of the difference

\be
\frac{\left< W(\gth)\right>-\left< W(0)\right>}{\left< W(0) \right>} = \frac{\left< \Tr\cP\,\e^{\oint \{A_\mu \dot x^\mu+\frac{2\pi}{k}M(0)C\bar C|\dot x|\}\dd\tau} \left[ \Tr\cP\,\e^{-\oint \frac{2\pi}{k}\, \frac{\gth^2}{2}\, M(0)C\bar C|\dot x|\}\dd\tau}  -1  \right] \right>}{\left< \Tr\cP\,\e^{\oint \{A_\mu \dot x^\mu+\frac{2\pi}{k}M(0)C\bar C|\dot x|\}\dd\tau} \right>}
\ee

It is related to the integral of two-point function of the composite scalar field $[MC\bar C](\tau)$ along the loop 

\be
\begin{split}
 =& - 2\gth^2 \left( \frac{2\pi}{k} \right)^2 \frac{N^2 \Gamma(1/2-\gep)^2}{\left( 4\pi^{\frac{3}{2}-\gep} \right)^2}  \int_0^{2\pi} \dd\tau_1\,\int_{\tau_1}^{2\pi} \dd\tau_2\, \frac{1}{[2(1-\cos(\tau_1-\tau_2))]^{1-2\gep}}\\
=& - \gth^2 \left( \frac{2\pi}{k} \right)^2 \frac{N^2 \Gamma(1/2-\gep)^2}{\left( 4\pi^{\frac{3}{2}-\gep} \right)^2}  \int_0^{2\pi} \dd\tau_1\,\int_0^{2\pi} \dd\tau_2\, \frac{1}{[2(1-\cos(\tau_1-\tau_2))]^{1-2\gep}}\\
\end{split}
\ee

In the last line we have used the symmetries of the integrand to symmetrise the domain of integration. One can now conformally map the integral in the last line above to the integral in (\ref{WWbar-pot}) by means of

\be
\frac{1}{2(1-\cos x)} \to \frac{1}{x^2} 
\ee

and conclude 

\be
 = -\frac{\gth^2}{4} \pi^2\gl^2 \frac{L^{4\gep}}{4\gep} +\cO(\gep)
\ee 

On the other hand expanding (\ref{non-pert-guess}) for small $\gth$ one has the simple relation

\be
W(\gl(1-\gth^2)) \sim W(\gl) - \gth^2\,\gl\,W(\gl)\frac{\d}{\d\gl}(t(\gl)+m(\gl,0))
\ee

We claim that this relation holds non-perturbatively in the 't~Hooft coupling $\gl$, supported by the results of sections \ref{fundamental-string}, \ref{weak-coupling-expansion} and \ref{strong-coupling-expansion} for the non-perturbative form of our Wilson loop. It then follows that the Bremsstrahlung function is given by

\be
\label{bremsstrahlung-function}
B(\gl) \sim \gl\frac{\d}{\d\gl}\log\left<W(0)\right> -\gl\frac{\d}{\d\gl}\log\left<W_{CS}\right>
\ee

where $W_{CS}$ is the Wilson loop in pure Chern-Simons theory.
The expectation value of the $\ov{6}$ BPS Wilson loop winding the equator of $S^3$ $n$ times, with $n\in\mathbb{N}$, was computed exactly in \cite{Klemm:2012ii} in terms of Airy functions. The standard loop operator $W(\gth =0)$ corresponds $n=1$.
Using the localization results of \cite{Klemm:2012ii} we can now compute the $B(\gl)$ functions relative to both the $\ov{6}$ and the $\half$ BPS Wilson loops. In the first case the logarithmic derivative reads

\begin{multline}
\label{d-W-ABJM-1-6}
N \d_N \log\left< W^{(1/6)} \right> = \\ 1-\pi\frac{N}{k}\cot\pi\frac{N}{k} +  N c \left[ \frac{c^2 (\gz-\gs) A_1(k) {\rm Ai}(c(\gz-\gs))-A_2(k){\rm Ai'}(c(\gz-\gs))}{c A_1(k) {\rm Ai'}(c(\gz-\gs)) - A_2(k) {\rm Ai}(c(\gz-\gs)) } - \frac{{\rm Ai'}(c\gz)}{{\rm Ai}(c\gz)} \right] 
\end{multline}

where 

\be
\begin{split}
& A_1(k)=\frac{\ii}{2\pi k}{\rm csc}\frac{2\pi}{k} \qquad A_2(k) =  \frac{2\pi}{k}{\rm csc}\frac{2\pi}{k} \left[ \left(\frac{k}{2}-\pi\cot\frac{2\pi}{k} \right) \frac{\ii}{2\pi^2} +\frac{k}{8\pi}\right]\\
& c= \left( \frac{\pi^2}{2}k \right)^{\ov{3}} \qquad\qquad \gz= N-\frac{k}{24}-\frac{1}{3k} \qquad\qquad \xi= \frac{6n}{3k}\\
\end{split}
\ee

Note that the first term is the contribution of pure Chern-Simons theory discussed in the Appendix (\ref{CS-shit}) in equation (\ref{awefull}) and following. In principle this result encodes all $1/N$ corrections to the $B$ function, but we don't expect that the equality holds beyond the 't~Hooft limit.  We can expand this result at arbitrary order in perturbation theory, we report the first few orders

\be
B^{(1/6)}(\gl) = 2 \pi ^2 \lambda ^2-\frac{3}{2} i \pi ^3 \lambda ^3-\frac{7
   \pi ^4 \lambda ^4}{3}+\frac{5}{2} i \pi ^5 \lambda
   ^5+\frac{241 \pi ^6 \lambda ^6}{60}-\frac{357}{80} i \pi
   ^7 \lambda ^7-\frac{18817 \pi ^8 \lambda
   ^8}{2520} + \cO\left(\lambda ^{9}\right)
\ee  

where the framing contribution has been taken apart. It is also straightforward to expand at large 't~Hooft coupling

\be
B^{(1/6)}(\gl) = \pi\sqrt{\frac{\gl}{2}} + \frac{3}{2}-\frac{2}{\pi} + \left( \frac{\ii}{8} + \ov{4\pi} + \frac{\pi}{96} \right)\frac{1}{\sqrt{2\gl}}  + \cO(\gl^{-1})
\ee

The result for the $\half$ BPS operator is more transparent

\be
B^{(1/2)}(\gl) = N\,c \left[ \frac{{\rm Ai'}(c(\gz-\xi))}{{\rm Ai}(c(\gz-\xi))} -\frac{{\rm Ai'}(c\gz)}{{\rm Ai}(c\gz)}  \right] - \pi\gl\cot \pi\gl+1
\ee

Again, one can expand $B(\gl)$ at high order for the singly-winding Wilson loop and strip the framing factor off, very carefully

\be
\begin{split}
& \gl\ll 1 \qquad B^{(1/2)}(\gl) = \pi^2 \gl^2 -\frac{2}{3}\pi^4 \gl^4 + \frac{17}{15}\pi^6 \gl^6 - \frac{5597}{2520}\pi^8 \gl^8 +\frac{481003}{45360}\pi^{10} \gl^{10} + \cO(\gl^{12}) \\
& \gl\bb 1 \qquad B^{(1/2)}(\gl) = \frac{\pi}{2} \sqrt{2\gl} + 1- \frac{2}{\pi} + \frac{\pi}{48\sqrt{2\gl}} + \cO(\gl^{-1})
\end{split}
\ee

The associated asymptotics for the cusp anomalous dimension agree at leading order with \cite{Griguolo:2012iq,Forini:2012bb} in both the weak and strong coupling regions and extend them in the non-perturbative region. The evaluation of the leading term in the strong coupling expansion is somewhat tricky. This term is actually $ 1-\pi\gl \,\cot\pi\gl$ and comes from the derivative acting on the Wilson loop (\ref{Wtopological}) of pure Chern-Simons theory in (\ref{bremsstrahlung-function}). It is known long since that the Chern-Simons level $k$ receives a one-loop correction that sets it to $k+N$ (being $N$ the rank of the group), and that this renormalization is non-perturbatively exact. For pure $SU(N)$ Chern-Simons theory the v.e.v. of Wilson loop is exactly known \cite{Witten:1988hf}.
In the large $\gl$ limit this function is perfectly well defined and one has (\ref{CS-shit})

\be
\gl \d_{\gl} \log\left< W_{\rm pure~CS} \right> = -1 +\cO(1/\gl)
\ee

On the other hand, although we can reorganise the perturbative expansion of ABJM theory in such a way to factorise the Chern-Simons contribution, in this case the level $k$ is protected by supersymmetry. The non-renormalised Chern-Simons Wilson loop (\ref{Wtopological}) displays an ill behaviour at strong coupling oscillating infinitely fast between $\pm\infty$. The best interpretation of this oscillatory behaviour that we were able to figure is to estimate its average value on a suitably wide interval, which turns out to be, for  sufficiently large $\gl$ (see Appendix \ref{CS-shit})

\be
\gl \d_{\gl} \log\left< W_{\rm CS} \right> = -1 +\frac{2}{\pi} +\cO(1/\gl)
\ee

\subsection{Bremsstrahlung and the magnon dispersion relation}

Superconformal Chern-Simons theory was first shown to be integrable at two-loop order in the $SU(4)$ sector in \cite{Minahan:2008hf,Minahan:2009te}. In \cite{Arutyunov:2008if,Stefanski:2008ik} the $AdS_4\times \mathbb{C}P^3$ sigma model was also shown to be integrable. The construction of the associated algebraic curve followed \cite{Gromov:2008bz} and a set of asymptotic Bethe equations interpolating from the latter result at strong coupling and two-loop Bethe ansatz of \cite{Minahan:2008hf} was proposed in \cite{Gromov:2008qe}. But from the investigation of the integrability properties of the $AdS_4/CFT_3$ correspondence a puzzling question emerges. Indeed, given $Q_n$ the set of all conserved charges, the spectrum of string's energies can be read from

\be
E= h(\gl) Q_2
\ee

The charges $Q_n$ are in turn given in term of the Bethe roots.
The function $h(\gl)$ was first introduced in \cite{Nishioka:2008gz,Gaiotto:2008cg,Grignani:2008is} and it enters the dispersion relation of a single magnon moving on the spin chain with momentum $p$ and $R-$charge $Q$ \cite{Beisert:2005tm}

\be
E(p) = \sqrt{Q^2 +4\, h^2(\gl) \,\sin^2 \frac{p}{2}} - Q
\ee

The common belief that $h^2(\gl)=\gl/(4\pi^2)$ in $AdS_5/CFT_4$ is consistent with results coming from both sides of the correspondence. On the other hand, in the context of $AdS_4/CFT_3$, this function is only known to interpolate between $h^2(\gl) =\gl^2 -4\gz_R(2)\,\gl^4 + \cO(\gl^6)$ for  $\gl\ll 1$ \cite{Minahan:2009aq} and $h(\gl) =\sqrt{\gl/2} + c_1 + \cO(1/\sqrt{\gl})$ for $\gl\bb 1$ \cite{Nishioka:2008gz,Gaiotto:2008cg,McLoughlin:2008he}. Here $\gz_R$ is the Riemann Zeta function. In a first instance it was proposed in \cite{Gromov:2008qe} to assign to $c_1$ the value zero for consistency of the TBA equations in the BMN limit. But this assumption actually disagreed with the computation of the energy of a closed spinning string in AdS \cite{Alday:2008ut}. Let us recall that, in $AdS_5\times S^5$, the relation between the energy $E$ and AdS spin $S$ of a closed spinning string that also carries $S^5$ angular momentum $J$ reads \cite{Frolov:2002av,Frolov:2006qe}

\be
E_{AdS_5}-S = \frac{1}{\pi} \left(\sqrt{\gl} - 3 \log\,2  \right)\,\log\frac{S}{J}
\ee 

in the limit where $J\ll S$. For twist-two operators, this relation actually defines the cusp anomalous dimension 

\be
\Delta-S= \Gamma(\gl)\log S \quad {\rm for} \quad S \to \infty \quad {\rm and} \quad  \frac{\phi^2}{2}\Gamma(\gl) = \lim_{\phi\to\ii\infty} \Gamma(\gl,\phi,0)
\ee

As a solution to the $h(\gl)$ puzzle, it was proposed in \cite{McLoughlin:2008he} that one should be able to map results obtained in $AdS_5\times S^5$ into  the corresponding ones in $AdS_3\times \mathbb{C}P^3$  by means of the substitution

\be
\frac{\sqrt{\gl_{\rm SYM}}}{4\pi} \to h(\gl_{\rm ABJM})=\sqrt{\frac{\gl_{\rm ABJM}}{2}}-\frac{\log 2}{2\pi}
\ee

In such a way the TBA equations correctly reproduce the BMN limit and the $AdS_5$ results of \cite{Frolov:2002av,Frolov:2006qe} are precisely mapped into their $AdS_4$ counterparts \cite{McLoughlin:2008ms,Alday:2008ut,Krishnan:2008zs}; in particular 

\be
E_{AdS_4}-S =  \left(2\,h(\gl) - \frac{3 \log\,2}{2\pi}  \right)\,\log\frac{S}{J} = 
\left(\sqrt{2\gl} - \frac{5 \log\,2}{2\pi}  \right)\,\log\frac{S}{J}
\ee 

Now we would like to compare the results obtained for the bremsstrahlung function $B(\gl)$, that is related to the expansion of $\Gamma_{\rm cusp}(\gl,\phi,\gth)$ near $\phi\sim\gth\sim 0$, to the function $h(\gl)$, that in turn is related to the expansion  $\Gamma_{\rm cusp}(\gl,\phi,\gth)$ in the limit $\phi,\gth\to\ii\infty$.
Quite interestingly we note that at the second non-trivial order at weak coupling

\be
B(\gl) = \pi^2(\gl^2-4\gl^4\gz_R(2)) = \pi^2 h^2(\gl)
\ee 

On the other hand at strong coupling the $h$ function behaves as

\be
\label{h-strong-coupling}
h(\gl) = \sqrt{\frac{\gl}{2}} - \frac{\log 2}{2\pi} + \cO\left(\ov{\sqrt{\gl}}\right) 
\ee

reason for which at leading order for $\gl\bb 1$ we still have

\be
B(\gl) \sim \pi \, h(\gl)
\ee

The coefficients of the $\gl^0$ terms differ in the two cases, though it appears that the two functions $B$ and $h$ are intimately related, even though they are related to the expansion of the cusp anomalous dimension in opposite kinematic regimes.

%%%%%%%%%%%%%%%%%%%%%%%%%%%%%%%%%%%%%%%%%%%%%%%%%%%%%%%%%%%%%%%%%%%%%%%%%%%%%%%%%%%

\section*{Comments}

We have constructed a new category of supersymmetric and purely bosonic Wilson loops for ABJM theory on $S^3$. These operators couple the scalar fields to a latitude on $S^2\in\mathbb{C}P^3$ at an angle $\gth$ and seem to be the three-dimensional analogue of the operators studied in \cite{Drukker:2006ga}. We argue that this operator corresponds in the theory localized on $S^3$ to the loop with winding number $\cos\gth$. We do not present a derivation of the localized form of the Wilson loop here, though we provide highly non-trivial checks at weak and strong coupling against a genuine field theoretic 2-loop computation and $AdS/CFT$ duality with type IIA superstrings on $AdS_4\times \mathbb{C}P^3$. In our case, these operators preserve $\ov{6}$ of the $\cN=6$ superconformal symmetry of ABJM theory, which is the least amount of supersymmetry on Euclidean $S^3$. It would be particularly interesting to generalise the construction of Section \ref{sec:weak} to super Wilson loops and introduce the couplings of fermionic fields along the lines of \cite{Cardinali:2012ru}. We suspect that in this way one should find the $\ov{3}$ BPS operators that are still missing in the ABJM zoo.\\

 Using the ABJM matrix model obtained in \cite{Kapustin:2009kz} through supersymmetric localization, and the subsequent exact solution of \cite{Marino:2011eh}, one can easily compute the expectation value of our operator at strong coupling and match the exact result in the limit $\gth\to 0$.
Availing on this result we were able to establish a connection between the cusp anomalous dimension and the logarithmic derivative of the Wilson loop. Quite interesting the relation differs from its four-dimensional analogue \cite{Correa:2012at} by an additional contribution of pure Chern-Simons theory. This contribution is in turn necessary to take into account the insensitivity of the CS field on the latitude angle $\gth$. This relation, which is perturbative in $\gth\sim 0$, defines non-perturbatively in the coupling constant the bremsstrahlung function $B(\gl)$ of $W$-bosons in three dimensions.\\

Further, it is known that the definition of the cusp anomalous dimension as the anomalous dimension of twist-two operators in the large spin limit creates a link between the expansion of $\Gamma(\gl,\phi,\gth)$ at small angles

$$  \Gamma(\gl,\phi,\gth)\sim (\phi^2-\gth^2)B(\gl) $$

and the analytic continuation to Minkowski space

$$  \Gamma(\gl,\phi,\gth)\sim \frac{\phi^2}{2} \Gamma_{\rm cusp}(\gl) $$

From the string point of view $\Gamma_{\rm cusp}$ is known \cite{McLoughlin:2008he,Krishnan:2008zs,Alday:2008ut} and it is believed to be

$$ \sim h(\gl) - \frac{3\log2}{2} $$

were $h(\gl)$ is the same function that enters into the TBA equations of \cite{Gromov:2008qe} and in the magnon dispersion relation \cite{Nishioka:2008gz,Gaiotto:2008cg,Grignani:2008is}. By comparing our result for $B(\gl)$ and the known expansions of $h(\gl)$ at weak and strong coupling we find a good agreement. Namely, the two-loop result of \cite{McLoughlin:2008he,Krishnan:2008zs,Alday:2008ut} is correctly matched as well as the leading strong coupling behaviour. \\

Though, we cannot reproduce the constant coefficient that appears in the strong coupling expansion of $h(\gl)$. Nonetheless the numerical value of our estimate is suspiciously close to the result of \cite{McLoughlin:2008he,Krishnan:2008zs,Alday:2008ut}. We suppose that this fact can be due to the roughness of our estimate and to the heavy assumptions that we have to take in order to treat the Chern-Simons contribution. In facts, the latter is singular whenever the coupling constant takes values $\gl=n\pi$ for any integer $n$. In this sense the point $\gl=\infty$ is somewhat special - it is the accumulation point of infinitely many singularities. From our conjecture it seems that the string theory in this regime actually senses the mean value of $\Gamma_{\rm cusp}$, obtained by averaging out these infinitely dense singularities. It would be of great interest to give a precise meaning to this last statement; but it might also be the case that $B$ and $h$ are different functions, since they appear in the expansion of the cusp anomaly in opposite kinematical regimes.

\section*{Acknowledgements}

I am thankful to Luca Griguolo, Domenico Seminara, Guido Festuccia and Konstantin Zarembo for fruitful discussions. Also, I acknowledge the Foundation Blanceflor Boncompagni-Ludovisi and Nordita for supporting my work.

%%%%%%%%%%%%%%%%%%%%%%%%%%%%%%%%%%%%%%%%%%%%%%%%%%%%%%%%%%%%%%%%%%%%%%%%%%%%%%

\appendix

\section{Conventions}
\label{sec:conventions}

We work in Euclidean signature where the rotated-time is in the third direction. As a basis for Gamma matrices we use the ordinary  Pauli matrices

\be
(\gamma^\mu)_\ga^{~\gb} = \{ \gs^1,\, \gs^2,\, \gs^3 \}
\ee

with

\be
\gs^1= \left(\begin{array}{rr}
0~ & 1 \\ 1~ & 0 \end{array}\right)
\qquad
\gs^2= \left(\begin{array}{rr}
0~ & -\ii \\ \ii~ & 0 \end{array}\right)
\qquad
\gs^3= \left(\begin{array}{rr}
1~ & 0 \\ 0~ & -1 \end{array}\right)
\ee

Spinor indices are risen and lowered with the antisymmetric tensor $\gep^{\ga\gb}$ using the convention $\gep^{12}=-\gep_{12}=1$. This implies

\be
\psi^\ga \xi_\ga =% - \xi_\ga \psi^\ga = 
\xi^\ga \psi_\ga \qquad
\psi^\ga (\gamma_\mu)_\ga^{~\gb} \xi_\gb = -\xi^\ga (\gamma_\mu)_\ga^{~\gb} \psi_\gb 
\qquad (\gamma^\mu)_{\ga\gb} = \{-\gs^3,\, \ii\mathbb{I},\, -\ii\gs^1 \}
\qquad (\gamma^\mu)^\ga_{~\gb}=-(\gamma^\mu)_\ga^{~\gb}
\ee

 Supersymmetry transformations $\gd_Q = \gth^\ga_{IJ} \bar Q_\ga^{IJ}$ are parametrised by $R-$antisymmetric spinors

\be
\label{eq:thetas-properties}
\gth_{IJ} = -\gth_{JI} \qquad \gth_{IJ} = \half \gep_{IJKL}\bar\gth^{KL} \qquad
(\gth_{IJ})^* = \bar\gth^{IJ}
\ee

and read

\be
\begin{split}
\label{susytrans}
& \gd A_\mu = \frac{4\pi\ii}{k} \left( \gth^\ga_{IJ} (\gamma_\mu)_\ga^{~\gb} \bar\psi^I_\gb \bar C^J + C_J \psi^\ga_I (\gamma_\mu)_\ga^{~\gb} \bar\gth_\gb^{IJ} \right)\\
& \gd C_I = 2\gth^\ga_{IJ} \bar\psi^J_\ga\\
& \gd \bar C^I = 2 \psi^\ga_J \bar\gth^{IJ}_\ga
\end{split}
\ee

Analogous equations hold for conformal transformations parametrised by $\gd_S=\eta^\ga_{IJ} \bar S_\ga^{IJ}$. The superconformal killing spinors have the form $\bar\vartheta^{IJ}_\ga = \bar\gth^{IJ}_\ga + (x\cdot\gamma)_\ga^{~\gb}\bar\eta^{IJ}_\gb$. Note that rising the spinor index will produce a flip in the sign of $\bar\vartheta_{IJ}^\ga = \bar\gth_{IJ}^\ga - \bar\eta_{IJ}^\gb (x\cdot\gamma)_\gb^{~\ga}$.

%%%%%%%%%%%%%%%%%%%%%%%%%%%%%%%%%%%%%%%%%%%%%%%%%%%%%%%%%%%%%%%%%%%%%%%%%%%%%%%%%%%%%%%

%\section{Killing spinors}
%\label{sec:killing spinors}

%%%%%%%%%%%%%%%%%%%%%%%%%%%%%%%%%%%%%%%%%%%%%%%%%%%%%%%%%%%%%%%%%%%%%%%%%%%%%%%%%%%%%%%%%%%%%%%%%%

\section{Chern-Simons contribution at strong coupling}
\label{CS-shit} % shit = strongly hybrid interacting theory ;-) 

The expectation value of the $SU(N)$ Chern-Simons Wilson loop was computed exactly in \cite{Witten:1988hf} 

\be
\left< W_{\rm pure~CS} \right> = \frac{\e^{\frac{\ii\pi}{k+N}\frac{N^2-1}{N}}}{N}\frac{\e^{\frac{\ii\pi N}{k+N}}-\e^{-\frac{\ii\pi N}{k+N}}}{\e^{\frac{\ii\pi}{k+N}}-\e^{-\frac{\ii\pi}{k+N}}}
\ee

Here the CS level $k$ appears shifted by the rank of the gauge group $N$, this is a well known one-loop renormalization of the coupling constant which is exact at higher loop orders. Acting with (\ref{bremsstrahlung-function}) one has

\be
\gl \frac{\d}{\d\gl} \log\left< W_{\rm pure~CS} \right> = 
N\frac{\d}{\d N}  \log\left< W_{\rm pure~CS} \right> = 
\frac{-(k + N)^2 +  N \pi \left( \cot\left( \frac{\pi}{k + N}\right) + k \cot\left(\frac{N \pi}{k + N}\right)\right)}{(k + N)^2}
\ee

by means of which the strong 't~Hooft coupling expansion simply reads 

\be
\gl \d_{\gl} \log\left< W_{\rm pure~CS} \right> = - 1 +\cO(1/\gl)
\ee

The situation is quite different when $k$ cannot receive any loop correction because of supersymmetry. In this case the Wilson loop becomes

\be
\label{CS-loop-no-shift}
\left< W \right> = \frac{\e^{-\ii\pi\gl}}{N} \frac{\sin\pi\gl}{\sin\frac{\pi}{k}}
\ee

Naively this function behaves as $\frac{1}{\gl}$ at large $\gl$, so should its logarithmic derivative. On the other hand (\ref{CS-loop-no-shift}) has an infinite number of zeros accumulating at $\gl=\infty$ which make the logarithmic derivative extremely ill-behaved 

\be
\label{awefull}
\gl \frac{\d}{\d\gl} \log\left< W \right> = -1 + \pi\gl \cot\pi\gl + \cO(1/\gl)
\ee

The ill contribution $\pi\gl \cot\pi\gl$ has an infinite number of poles accumulating at $\gl=\infty$. We want to estimate the average of this term over a sufficiently large interval in the strongly coupled region. To this aim we consider the integral

\be
I = \lim_{L\to\infty}\frac{1}{L} \int_L^{2L} \dd x \, \e^{x\,\cot x} = \lim_{L\to\infty}\frac{1}{L} \sum_n \e^{x_n \cot x_n}
\ee

In the interval $[L,2L]$ it has $\frac{L}{\pi}$ poles located at $x_n=n\pi$. For $x$ approaching $x_n$ from the left the function behaves as

\be
I = \lim_{\gep\to 0} (x_n - \gep)\cot(x_n - \gep) = (n\pi-\gep)\left(-\ov{\gep}+\frac{\gep}{3} \right) + \cO(\gep^2)
\ee

Then we can write the sum as

\be
I = \lim_{\gep\to 0} \lim_{L\to\infty}\frac{1}{L} \sum_{n=[L\pi]}^{[2L/\pi]} \left[ (n\pi-\gep)\left(-\ov{\gep}+\frac{\gep}{3} \right) +(n\pi+\gep)\left(+\ov{\gep}-\frac{\gep}{3} \right) \right] =  \lim_{\gep\to 0} \frac{2}{\pi}\left( 1-\frac{\gep^2}{3} \right)
\ee

where $[L/\pi]$ means the integer part.  We are then lead to conclude that the strong coupling average of (\ref{awefull}) is finite and reads

\be
\bigg< \gl \frac{\d}{\d\gl} \log\left< W \right>\, \bigg> = -1 + \frac{2}{\pi}
\ee

%%%%%%%%%%%%%%%%%%%%%%%%%%%%%%%%%%%%%%%%%%%%%%%%%%%%%%%%%%%%%%%%%%%%%%%%%%%%%%%%%%%%%%%

\section{Matrix model results for the ABJM Wilson loop}
\label{sec:CS-localization}

The Wilson loop in $SU(N_1)\times SU(N_2)$ superconformal Chern-Simons theories coupled to matter  on $S^3$ has been localized to a matrix integral in \cite{Kapustin:2009kz} using the procedure introduced in \cite{Pestun:2007rz} for four-dimensional SYM theories on $S^4$ with 2 and 4 supersymmetries. The partition function is computed by an integral with a still quadratic potential, but with a highly non-trivial measure, opposed to the $\cN=4$ four-dimensional case where the integral is Gaussian and explicitly solvable. As it was shown in \cite{Kapustin:2009kz}, the vacuum expectation value of the {\small $\frac{1}{6}$}BPS Wilson loop of \cite{Drukker:2008zx} in a representation $\cR$ of either $SU(N_1)$ at level $k$ or $SU(N_2)$ at level $-k$ is computed by the quantum average

\be
\left< W^{1/6}_\cR \right> = \frac{1}{\dim\cR} \left< \tr_\cR \e^{\mu_i} \right>_{\small{ABJM}}
\ee  

with respect to 

\be
\label{eq:ZABJMloc}
Z_{\small{ABJM}}(N_1,N_2,g) = \int \prod_{i=1}^{N_1} \dd\mu_i\, \prod_{j=1}^{N_2} \dd\nu_j \,
\frac{\prod_{i<j} \sinh^2\left( \frac{\mu_i-\mu_j}{2} \right)\,\sinh^2\left( \frac{\nu_i-\nu_j}{2} \right) }{\prod_{i,j}\cosh^2\left( \frac{\mu_i-\nu_j}{2} \right)} \, \e^{-\ov{2g}\left[ \sum_i \mu_i^2 - \sum_j \nu_j^2 \right]} 
\ee

It should be emphasised that in the $\cN=2$ language, the natural supersymmetric Wilson loop couples to the auxiliary scalar $\gs$ of the vector multiplet as

\be
\label{eq:1/6operator}
\left< W_\cR \right> = \left< \tr_\cR\,\cP\, \e^{\oint A_\mu\dd x^\mu +\gs|\dot x|}\right>
\ee

which is also natural as the localization locus is given by the configurations in which all fields vanish except $\gs$, that in turn takes constant values. On the other hand, once the equations of motion are taken into account one finds
\cite{Benna:2008zy}

\be
\label{eq:sigma-in-terrms-of-C}
\gs^a = \ov{4k} \tr_\cR T^a \left(C_1^2+C_2^2 - C_3^2 -C_4^2 \right)
\ee

being $C_I^2= C_I \bar C^I$, $T^a \in \mathfrak{su}(N)$, that immediately yields to the operator considered in \cite{Drukker:2008zx} after  $\gs$ has been integrated out.

In addition,  $SU(N_1)\times SU(N_2)$ Chern-Simons theories with $\cN\geq 4$ supersymmetry display the presence of an underlying $SU(N_1|N_2)$ supergroup structure \cite{Gaiotto:2008sd}. This fact originally motivated the proposal of a ${\small \half}$BPS operator based on the quantum holonomy of a $SU(N_1|N_2)$ superconnection \cite{Drukker:2009hy}. Based on this consideration, in \cite{Drukker:2009hy} the localized matrix integral for the ${\small \half}$BPS loop in a super-representation $\cS$ of  $SU(N_1|N_2)$ was proposed to be given by

\be
\left< W_{\cS} \right> = \ov{\dim \cS} \, \left< \rm{Str} \left( \begin{array}{cc}
\e^{\mu_i} & 0 \\ 0 & -\e^{\nu_i} \end{array} \right) \right>_{\small ABJM}
\ee

where ${\rm Str}$ is the supertrace in $\cS$. The integral above was then recognised \cite{Marino:2009jd}  to be the supergroup extension of the correlator 

\be
\label{eq:correinlensspace}
\left< W_{\cR} \right> = \ov{\dim \cR} \, \left< \rm{tr} \left( \begin{array}{cc}
\e^{\mu_i} & 0 \\ 0 & -\e^{\nu_i} \end{array} \right) \right>_{\small ABJM}
\ee

 in a representation $\cR$ of $SU(N_1+N_2)$ in the already known two-cut matrix model describing Chern-Simons theory on the lens space $L(2,1)$
\footnote{Three dimensional lens spaces are defined as the quotients $L(p,1)= S^3 /\mathbb{Z}_p$. The $1/N$ expansion of CS gauge theory on $L(2,1)$ coincides with the genus expansion of topological string theory on the local Calabi-Yau manifold given by the anti-canonical bundle of the Hirzebruch surface $\mathbb{C}P^1 \times \mathbb{C}P^1$ \cite{Aganagic:2002wv}. The quantum correlator in (\ref{eq:correinlensspace}) then has also a Gopakumar-Vafa dual interpretation in terms of an open topological string amplitude \cite{Aganagic:2002wv}. It is interesting to note that from the point of view of topological string theory on $\mathbb{C}P^1 \times \mathbb{C}P^1$, the more natural object is then the supercorrelator, or ${\small \half}$BPS Wilson loop, rather then the ${\small \ov{6}}$BPS one. The lens space matrix model was first introduced in \cite{Marino:2002fk}.}, 
up to the analytical continuation 

\be
N_2 \to -N_2
\ee

that is in facts equivalent to changing sign to one of the (rescaled) 't~Hooft couplings 

\be
t_1 = 2\pi\ii\frac{N_1}{k} \qquad t_2= -2\pi\ii\frac{N_2}{k}
\ee

The lens space matrix model was deeply investigated in \cite{Marino:2009jd,Drukker:2010nc,Drukker:2011zy,Fuji:2011km,Marino:2011eh,Klemm:2012ii} using standard matrix models technology and also building on known results \cite{Halmagyi:2003ze, Halmagyi:2003mm}. 

When the super-representation $\cS$ of $SU(N_1|N_2)$ is the one induced by the adjoint representation of $SU(N_1+N_2)$, one has a simple relation between the two supersymmetric loops. Indeed, setting $W_{\rm adj}$ the operator in (\ref{eq:1/6operator}) for the gauge field $A_\mu$ in the adjoint representation of $SU(N_1)$ and $\widehat{W}_{\rm adj}$ its homologous for the gauge field $\hat A_\mu$ in the adjoint of $SU(N_2)$ one has

\be
\label{eq:relation1/2-1/6}
\left< W_{\cS} \right> = \frac{1}{N_1 - N_2} \left(N_1 \big< W_{\rm adj}\big> - N_2 \big< \widehat{W}_{\rm adj}\big>   \right)
\ee

This gives a hint that the behaviour of the two observables at strong coupling must be the same, at least to leading order, in the ABJM theory. Indeed the symmetry of the action allows for differences between the perturbative expansions of $W_{\rm adj}$ and $\widehat{W}_{\rm adj}$, in the planar limit, only at odd orders of $N_1,\, N_2$ respectively, which are then canceled by the ABJM projection ($N_1=N_2$). Also note that a minus sign difference persists between the two  {\small$\ov{6}$}BPS operators due to the fact that one field has Chern-Simons level $k$ while the other has level $-k$. Moreover, (\ref{eq:relation1/2-1/6}) is perfectly well defined for $N_1=N_2$.\\

In the large $N$ limit the density of eigenvalues encoding the master field solution of the ABJM matrix model was computed in \cite{Marino:2009jd} and reads

\be
\label{eigen-density}
\rho_1(x)\dd x = \frac{1}{\pi t_1} \tan^{-1} \sqrt{\frac{\ga x-1-x^2}{\gb x+1+x^2}}\, \frac{\dd x}{x}
\qquad
\rho_2(x)\dd x = -\frac{1}{\pi t_2} \tan^{-1} \sqrt{\frac{\gb x+1+x^2}{\ga x-1-x^2}}\, \frac{\dd x}{x}
\ee

where $\ga,\gb$ are related to the endpoints of the two cuts and read in terms of the mirror coupling constant

\be
\ga=2+\ii\kappa \qquad\qquad \gb=2 -\ii\kappa
\ee

 In turn the mirror coupling is related to the 't~Hooft coupling through the mirror map

\be
\gl(\kappa)= \frac{\kappa}{8\pi} ~_3F_2\left(\half,\half,\half;1,\frac{3}{2};-\frac{\kappa^2}{16}  \right)
\ee

In these settings the $\ov{6}$ BPS Wilson loop of winding number $n$ is given by the integral

\be
\label{eq:loop-winding-n}
\left< W(\gth)\right> = \int\dd\mu \rho(\mu)\e^{n\,\mu}
\ee

with support over the interval $[1/a,\,a]$. The Wilson loop of the second gauge group has support on the interval $[1/b,\,b]$ and density $\rho_2(x)$. The endpoints of the cuts are also given in terms of the mirror map

\be
\begin{split}
a(k) &=\half\left( 2+\ii \kappa+\sqrt{\kappa(4\ii-\kappa)} \right)\\
b(k) &=\half\left( 2-\ii \kappa+\sqrt{-\kappa(4\ii+\kappa)} \right)
\end{split}
\ee

Integrating explicitly the eigenvalue density is a hard task, but its derivatives with respect to $\kappa$ can be integrated exactly. This allows to write a differential equation for the Wilson loop itself

\be
\frac{\d}{\d\kappa}\left( \gl(\kappa)\,W(\kappa) \right) = \frac{\d}{\d\kappa}\int_{1/a}^a x^{\ii\pi n }\rho(x)\dd x
\ee

that can be solved order by order in $\kappa\ll 1$ and then inverted to get $W$ as a function of $\gl$. 
This enabled the authors of  \cite{Marino:2009jd} to compute the expansion of both the $\ov{6}$ and the $\half$BPS Wilson loops at weak and strong coupling and match previous results coming from gauge theory and semiclassical string computations \cite{Drukker:2008zx,Rey:1998ik}. Subsequently a reformulation of the ABJM matrix model in terms of Fermi gas was proposed \cite{Marino:2011eh}. This approach allowed the computation of the partition function and later on of the supersymmetric Wilson loop at all orders in the coupling (meaning only up to instanton contributions that are exponentially suppressed in the large $N$ limit) \cite{Klemm:2012ii}

\be
\left< W^{(1/2)} \right> = \frac{1}{4}{\rm csc}\frac{2\pi n}{k}\left[\frac{{\rm Ai}(c(\gz-\xi))}{{\rm Ai}(c\gz)}  \right]
\ee

\be
c= \left( \frac{\pi^2}{2}k \right)^{\ov{3}} \qquad\qquad \gz= N-\frac{k}{24}-\frac{1}{3k} \qquad\qquad \xi= \frac{6n}{3k}
\ee

Here $n$ is again the winding number of the loop, the number of times it encircles the equator of $S^3$. Note that the latter result encodes all $1/N$  corrections in terms of a genus expansion of the matrix model.

%%%%%%%%%%%%%%%%%%%%%%%%%%%%%%%%%%%%%%%%%%%%%%%%%%%%%%%%%%%%%%

%%%%%%%%%%%%%%%%%%%%%%%%%%%%%%%%%%%%%%%%%%%%%%%%%%%%%%%%%%%%%%

\newpage
\bibliographystyle{JHEP}
\bibliography{biblio}

\end{document}